\documentclass[journal,comsoc]{IEEEtran}

\usepackage[utf8]{inputenc}
\usepackage[english]{babel}
\usepackage[T1]{fontenc}

\usepackage{cite}

   \usepackage[pdftex]{graphicx}
   \DeclareGraphicsExtensions{.pdf,.jpeg,.png}
   \usepackage{everysel}
   \usepackage{keyval}
   \usepackage{ragged2e}

\usepackage{epstopdf}

\usepackage{amsmath}
\usepackage{mathtools}

\usepackage{textcomp}
\usepackage{siunitx}
\usepackage{gensymb}
\usepackage{balance}
\usepackage{bm}
\usepackage{xr}
\usepackage{array}
\usepackage{pbox}
\usepackage{booktabs}
\usepackage{xcolor}

\usepackage[font=footnotesize]{caption}
\usepackage[font=footnotesize]{subcaption}

\usepackage{soul}

\usepackage{colortbl}
\newcommand{\new}[1]{\textcolor{black}{{#1}}}
\newcommand{\newrev}[1]{\textcolor{black}{{#1}}}

\usepackage{algorithmic}

\usepackage{float}

\usepackage[acronyms,nonumberlist,nopostdot,nomain,nogroupskip]{glossaries}

\newacronym{rfi}{RFI}{Radio Frequency Interference}
\newacronym{lb}{LB}{Lower Band}
\newacronym{ub}{UB}{Upper Band}
\newacronym{awg}{AWG}{Arbitrary Waveform Generator}
\newacronym{dso}{DSO}{Digital Storage Oscilloscope}
\newacronym{3gpp}{3GPP}{3rd Generation Partnership Project}
\newacronym{adc}{ADC}{Analog to Digital Converter}
\newacronym{5g}{5G}{5th Generation}
\newacronym{6g}{6G}{6th Generation}
\newacronym{aimd}{AIMD}{Additive Increase Multiplicative Decrease}
\newacronym{am}{AM}{Acknowledged Mode}
\newacronym{amc}{AMC}{Adaptive Modulation and Coding}
\newacronym{aqm}{AQM}{Active Queue Management}
\newacronym{awgn}{AGWN}{Additive White Gaussian Noise}
\newacronym{balia}{BALIA}{Balanced Link Adaptation}
\newacronym{bdp}{BDP}{Bandwidth-Delay Product}
\newacronym{bf}{BF}{Beamforming}
\newacronym{cc}{CC}{Congestion Control}
\newacronym{cdf}{CDF}{Cumulative Distribution Function}
\newacronym{cn}{CN}{Core Network}
\newacronym{cqi}{CQI}{Channel Quality Information}
\newacronym{cp}{CP}{Control Plane}
\newacronym{csirs}{CSI-RS}{Channel State Information - Reference Signal}
\newacronym{dc}{DC}{Dual Connectivity}
\newacronym{dce}{DCE}{Direct Code Execution}
\newacronym{dci}{DCI}{Downlink Control Information}
\newacronym{dl}{DL}{Downlink}
\newacronym{dmr}{DMR}{Deadline Miss Ratio}
\newacronym{dmrs}{DMRS}{DeModulation Reference Signal}
\newacronym{e2e}{E2E}{End-to-End}
\newacronym{ecn}{ECN}{Explicit Congestion Notification}
\newacronym{edf}{EDF}{Earliest Deadline First}
\newacronym{enb}{eNB}{evolved Node Base}
\newacronym{epc}{EPC}{Evolved Packet Core}
\newacronym{es}{ES}{Edge Server}
\newacronym{fdma}{FDMA}{Frequency Division Multiple Access}
\newacronym{fdd}{FDD}{Frequency Division Duplexing}
\newacronym{fft}{FFT}{Fast Fourier Transform}
\newacronym[firstplural=Radio Access Technologies (RATs)]{rat}{RAT}{Radio Access Technology}
\newacronym{fs}{FS}{Fast Switching}
\newacronym{ftp}{FTP}{File Transfer Protocol}
\newacronym{gnb}{gNB}{Next Generation Node Base}
\newacronym{harq}{HARQ}{Hybrid Automatic Repeat reQuest}
\newacronym{hetnet}{HetNet}{Heterogeneous Network}
\newacronym{hh}{HH}{Hard Handover}
\newacronym{hol}{HOL}{Head-of-Line}
\newacronym{ia}{IA}{Initial Access}
\newacronym{imt}{IMT}{International Mobile Telecommunication}
\newacronym{iot}{IoT}{Internet of Things}
\newacronym{los}{LOS}{Line-of-Sight}
\newacronym{lte}{LTE}{Long Term Evolution}
\newacronym{m2m}{M2M}{Machine to Machine}
\newacronym{mac}{MAC}{Medium Access Control}
\newacronym{mc}{MC}{Multi-Connectivity}
\newacronym{mcs}{MCS}{Modulation and Coding Scheme}
\newacronym{mec}{MEC}{Mobile Edge Cloud}
\newacronym{mi}{MI}{Mutual Information}
\newacronym{mimo}{MIMO}{Multiple Input, Multiple Output}
\newacronym{mmwave}{mmWave}{millimeter wave}
\newacronym{mptcp}{MPTCP}{Multipath TCP}
\newacronym{mr}{MR}{Maximum Rate}
\newacronym{mss}{MSS}{Maximum Segment Size}
\newacronym{mtd}{MTD}{Machine-Type Device}
\newacronym{mtu}{MTU}{Maximum Transmission Unit}
\newacronym{nfv}{NFV}{Network Function Virtualization}
\newacronym{nlos}{NLOS}{Non-Line-of-Sight}
\newacronym{nr}{NR}{New Radio}
\newacronym{ofdm}{OFDM}{Orthogonal Frequency Division Multiplexing}
\newacronym{papr}{PAPR}{Peak-to-Average Power Ratio}
\newacronym{pdcch}{PDCCH}{Physical Downlonk Control Channel}
\newacronym{pdcp}{PDCP}{Packet Data Convergence Protocol}
\newacronym{pdsch}{PDSCH}{Physical Downlink Shared Channel}
\newacronym{pdu}{PDU}{Packet Data Unit}
\newacronym{pf}{PF}{Proportional Fair}
\newacronym{pgw}{PGW}{Packet Gateway}
\newacronym{phy}{PHY}{physical}
\newacronym{pbch}{PBCH}{Physical Broadcast Channel}
\newacronym[plural=\gls{mme}s,firstplural=Mobility Management Entities (MMEs)]{mme}{MME}{Mobility Management Entity}
\newacronym{prb}{PRB}{Physical Resource Block}
\newacronym{pss}{PSS}{Primary Synchronization Signal}
\newacronym{pucch}{PUCCH}{Physical Uplink Control Channel}
\newacronym{pusch}{PUSCH}{Physical Uplink Shared Channel}
\newacronym{rach}{RACH}{Random Access Channel}
\newacronym{ran}{RAN}{Radio Access Network}
\newacronym{red}{RED}{Random Early Detection}
\newacronym{rf}{RF}{Radio Frequency}
\newacronym{rlc}{RLC}{Radio Link Control}
\newacronym{rlf}{RLF}{Radio Link Failure}
\newacronym{rrc}{RRC}{Radio Resource Control}
\newacronym{rrm}{RRM}{Radio Resource Management}
\newacronym{rr}{RR}{Round Robin}
\newacronym{rs}{RS}{Remote Server}
\newacronym{rsrp}{RSRP}{Reference Signal Received Power}
\newacronym{rss}{RSS}{Received Signal Strength}
\newacronym{rtt}{RTT}{Round Trip Time}
\newacronym{rw}{RW}{Receive Window}
\newacronym{rx}{RX}{Receiver}
\newacronym{sa}{SA}{standalone}
\newacronym{sack}{SACK}{Selective Acknowledgment}
\newacronym{sap}{SAP}{Service Access Point}
\newacronym{sch}{SCH}{Secondary Cell Handover}
\newacronym{scoot}{SCOOT}{Split Cycle Offset Optimization Technique}
\newacronym{sdma}{SDMA}{Spatial Division Multiple Access}
\newacronym{sinr}{SINR}{Signal to Interference plus Noise Ratio}
\newacronym{sm}{SM}{Saturation Mode}
\newacronym{snr}{SNR}{Signal-to-Noise-Ratio}
\newacronym{son}{SON}{Self-Organizing Network}
\newacronym{ss}{SS}{Synchronization Signal}
\newacronym{srs}{SRS}{Sounding Reference Signal}
\newacronym{sss}{SSS}{Secondary Synchronization Signal}
\newacronym{tb}{TB}{Transport Block}
\newacronym{tcp}{TCP}{Transmission Control Protocol}
\newacronym{tdd}{TDD}{Time Division Duplexing}
\newacronym{tdma}{TDMA}{Time Division Multiple Access}
\newacronym{tfl}{TfL}{Transport for London}
\newacronym{tm}{TM}{Transparent Mode}
\newacronym{trp}{TRP}{Transmitter Receiver Pair}
\newacronym{tti}{TTI}{Transmission Time Interval}
\newacronym{ttt}{TTT}{Time-to-Trigger}
\newacronym{tx}{TX}{Transmitter}
\newacronym{ue}{UE}{User Equipment}
\newacronym{ul}{UL}{Uplink}
\newacronym{uml}{UML}{Unified Modeling Language}
\newacronym{um}{UM}{Unacknowledged Mode}
\newacronym{utc}{UTC}{Urban Traffic Control}
\newacronym{vm}{VM}{Virtual Machine}
\newacronym{rsrq}{RSRQ}{Reference Signal Received Quality}
\newacronym{rssi}{RSSI}{Received Signal Strength Indicator}
\newacronym{crs}{CRS}{Cell Reference Signal}
\newacronym{nsa}{NSA}{Non Stand Alone}
\newacronym{mrdc}{MR-DC}{Multi \gls{rat} \gls{dc}}
\newacronym{endc}{EN-DC}{E-UTRAN-\gls{nr} \gls{dc}}
\newacronym{5gc}{5GC}{5G Core}
\newacronym{si}{SI}{Study Item}
\newacronym{iab}{IAB}{Integrated Access and Backhaul}
\newacronym{wf}{WF}{Wired-first}
\newacronym{hqf}{HQF}{Highest-quality-first}
\newacronym{pa}{PA}{Position-aware}
\newacronym{mlr}{MLR}{Maximum-local-rate}
\newacronym{wbf}{WBF}{Wired Bias Function}
\newacronym{mib}{MIB}{Master Information Block}
\newacronym{sib}{SIB}{Secondary Information Block}
\newacronym{kpi}{KPI}{Key Performance Indicator}
\newacronym{ppp}{PPP}{Poisson Point Process}
\newacronym{gtp}{GTP}{GPRS Tunneling Protocol}
\newacronym{amf}{AMF}{Access and Mobility Management Function}
\newacronym{dash}{DASH}{Dynamic Adaptive Streaming over HTTP}
\newacronym{http}{HTTP}{HyperText Transfer Protocol}
\newacronym{qos}{QoS}{Quality of Service}
\newacronym{udp}{UDP}{User Datagram Protocol}
\newacronym{cu}{CU}{Central Unit}
\newacronym{du}{DU}{Distributed Unit}
\newacronym{mt}{MT}{Mobile Termination}
\newacronym{sdap}{SDAP}{Service Data Adaptation Protocol}
\newacronym{tdm}{TDM}{Time Division Multiplexing}
\newacronym{fdm}{FDM}{Frequency Division Multiplexing}
\newacronym{sdm}{SDM}{Space Division Multiplexing}
\newacronym{dag}{DAG}{Directed Acyclic Graph}
\newacronym{st}{ST}{Spanning Tree}
\newacronym{uav}{UAV}{Unmanned Aerial Vehicle}
\newacronym{ber}{BER}{Bit Error Rate}
\newacronym{fcc}{FCC}{Federal Communications Commission}
\newacronym{itu}{ITU}{International Telecommunications Union}
\newacronym{spdt}{SPDT}{Single-Pole Double Throw}
\newacronym{mls}{MLS}{Microwave Limb Scanner}
\newacronym{api}{API}{Application Programming Interface}
\newacronym{ntp}{NTP}{Network Time Protocol}
\newacronym{rest}{REST}{Representational state transfer}
\newacronym{lo}{LO}{Local Oscillator}
\newacronym{if}{IF}{Intermediate Frequency}
\newacronym{qam}{QAM}{Quadrature Amplitude Modulation}
\newacronym{ummimo}{UM-MIMO}{Ultra Massive \gls{mimo}}
\newacronym{cbrs}{CBRS}{Citizen Broadband Radio Service}
\newacronym{noma}{NOMA}{Non-Orthogonal Multiple Access}
\newacronym{rre}{RR}{Radio Regulations}
\newacronym{wrc}{WRC}{World Radio Conference}
\newacronym{cept}{CEPT}{European Conference of Postal and Telecommunications Administrations}
\newacronym{ehf}{EHF}{Extremely High Frequencies}
\newacronym{uhf}{UHF}{Ultra High Frequencies}
\newacronym{shf}{SHF}{Super High Frequencies}
\newacronym{ras}{RAS}{Radio Astronomy Service}
\newacronym{eess}{EESS}{Earth-Exploration Satellite Service}
\newacronym{ses}{SES}{Space-Exploration Service}
\newacronym{dsss}{DSSS}{Direct Sequence Spread Spectrum}
\newacronym{cdma}{CDMA}{Code Division Multiple Access}
\newacronym{sll}{SLL}{side-lobe-levels}
\newacronym{bs}{BS}{base station}
\newacronym{psd}{PSD}{Power Spectral Density}
\newacronym{fss}{FSS}{Frequency Selective Surface}
\newacronym{umts}{UMTS}{Universal Mobile Telecommunications System}
\newacronym{nrdz}{NRDZ}{National Radio Dynamic Zone}

\usepackage{tikz}
\usepackage{pgfplots}
\pgfplotsset{compat=newest}
\pgfplotsset{plot coordinates/math parser=false}
\newlength\fheight
\newlength\fwidth
\usetikzlibrary{plotmarks,patterns,decorations.pathreplacing,backgrounds,calc,arrows,arrows.meta,spy,matrix}
\usepgfplotslibrary{patchplots,groupplots}
\usepackage{tikzscale}

\tikzstyle{startstop} = [rectangle, rounded corners, minimum width=2cm, minimum height=0.5cm,text centered, draw=black]
\tikzstyle{io} = [trapezium, trapezium left angle=70, trapezium right angle=110, minimum width=3cm, minimum height=1cm, text centered, draw=black]
\tikzstyle{process} = [rectangle, minimum width=2cm, minimum height=0.5cm, text centered, draw=black, alignb=center]
\tikzstyle{decision} = [ellipse, minimum width=2cm, minimum height=1cm, text centered, draw=black]
\tikzstyle{arrow} = [thick,<->,>=stealth]
\tikzstyle{line} = [thick,>=stealth]
\tikzstyle{darrow} = [thick,<->,>=stealth,dashed]
\tikzstyle{sarrow} = [thick,->,>=stealth]
\tikzstyle{larrow} = [line width=0.1mm,dashdotted,->,>=stealth]

\makeatletter
\def\grd@save@target#1{%
  \def\grd@target{#1}}
\def\grd@save@start#1{%
  \def\grd@start{#1}}
\tikzset{
  grid with coordinates/.style={
    to path={%
      \pgfextra{%
        \edef\grd@@target{(\tikztotarget)}%
        \tikz@scan@one@point\grd@save@target\grd@@target\relax
        \edef\grd@@start{(\tikztostart)}%
        \tikz@scan@one@point\grd@save@start\grd@@start\relax
        \draw[minor help lines] (\tikztostart) grid (\tikztotarget);
        \draw[major help lines] (\tikztostart) grid (\tikztotarget);
        \grd@start
        \pgfmathsetmacro{\grd@xa}{\the\pgf@x/1cm}
        \pgfmathsetmacro{\grd@ya}{\the\pgf@y/1cm}
        \grd@target
        \pgfmathsetmacro{\grd@xb}{\the\pgf@x/1cm}
        \pgfmathsetmacro{\grd@yb}{\the\pgf@y/1cm}
        \pgfmathsetmacro{\grd@xc}{\grd@xa + \pgfkeysvalueof{/tikz/grid with coordinates/major step x}}
        \pgfmathsetmacro{\grd@yc}{\grd@ya + \pgfkeysvalueof{/tikz/grid with coordinates/major step y}}
        \foreach \x in {\grd@xa,\grd@xc,...,\grd@xb}
        \node[anchor=north] at (\x,\grd@ya) {\pgfmathprintnumber{\x}};
        \foreach \y in {\grd@ya,\grd@yc,...,\grd@yb}
        \node[anchor=east] at (\grd@xa,\y) {\pgfmathprintnumber{\y}};
      }
    }
  },
  minor help lines/.style={
    help lines,
    gray,
    line cap =round,
    xstep=\pgfkeysvalueof{/tikz/grid with coordinates/minor step x},
    ystep=\pgfkeysvalueof{/tikz/grid with coordinates/minor step y}
  },
  major help lines/.style={
    help lines,
    line cap =round,
    line width=\pgfkeysvalueof{/tikz/grid with coordinates/major line width},
    xstep=\pgfkeysvalueof{/tikz/grid with coordinates/major step x},
    ystep=\pgfkeysvalueof{/tikz/grid with coordinates/major step y}
  },
  grid with coordinates/.cd,
  minor step x/.initial=.5,
  minor step y/.initial=.2,
  major step x/.initial=1,
  major step y/.initial=1,
  major line width/.initial=1pt,
}
\makeatother

\newif\ifexttikz
\exttikzfalse

\ifexttikz
  \usetikzlibrary{external}
\fi

\usepackage{dblfloatfix}    

\usepackage{multicol}
\usepackage{tabularx}

\begin{document}
%

\title{Coexistence and Spectrum Sharing Above 100~GHz}

\author{Michele Polese,~\IEEEmembership{Member,~IEEE}, Xavier Cantos-Roman,~\IEEEmembership{Graduate Student Member,~IEEE},\\Arjun Singh,~\IEEEmembership{Member,~IEEE}, Michael J. Marcus,~\IEEEmembership{Fellow,~IEEE}, Thomas J. Maccarone,\\Tommaso Melodia,~\IEEEmembership{Fellow,~IEEE}, and Josep~M.~Jornet,~\IEEEmembership{Senior Member,~IEEE}%

\thanks{M. Polese, X. Cantos-Roman, M. J. Marcus, T. Melodia, and J. M. Jornet are with the Institute for the Wireless Internet of Things and with the Department of Electrical and Computer Engineering, Northeastern University, Boston, MA, USA.}%
\thanks{A. Singh is with the Department of Engineering at the State University of New York Polytechnic Institute, Utica, NY, USA.}
\thanks{T. J. Maccarone is with the Department of Physics \& Astronomy, Texas Tech University, Lubbock, TX, USA.}%
\thanks{This work was partially supported by NSF under Grants AST-2037896, MPS-2037890, CNS-2011411, and CNS-2225590.}
}

\bstctlcite{BSTcontrol}  

\maketitle

\begin{abstract}

The electromagnetic spectrum plays a fundamental role for the development of the digital society. It enables wireless communications (either between humans or machines) and sensing (for example for Earth exploration, radio astronomy, imaging, and radars, among others). While each of these uses benefits from a larger bandwidth, the spectrum is a finite resource. This introduces competing interests among the different stakeholders of the spectrum, which have led---so far---to rigid policies and spectrum allocations. Recently, the spectrum crunch in the sub-6 GHz bands has prompted communication technologies to move to higher carrier frequencies, where future 6G wireless networks can exploit theoretically very large bandwidths. However, the spectrum above 100~GHz features several narrow, yet numerous sub-bands that are exclusively allocated for passive sensing applications, e.g., for climate and weather monitoring. This prevents the allocation of large contiguous bands to active users of the spectrum, either being communications (which need tens of gigahertz of bandwidth to target terabit-per-second links) or radars. 
This paper explores how spectrum policy and spectrum technologies can evolve to enable \textit{sharing} among different stakeholders in the above 100~GHz spectrum, without introducing harmful interference or disrupting either security applications or fundamental science exploration. \new{This portion of the spectrum presents new challenges and opportunities for the design of spectrum sharing schemes, including higher spreading and absorption losses, extremely directional antenna technologies, and ultra-high data-rate communications, among others.}
The paper provides a tutorial on current regulations above 100~GHz, and highlights how sharing is central to allowing each stakeholder to make the most out of this spectrum. 
It then defines---through detailed simulations based on standard \gls{itu} channel and antenna models---scenarios in which active users may introduce harmful interference to passive sensing. Based on this evaluation, it reviews a number of promising techniques that can enable active/passive sharing above 100~GHz. 
The critical review and tutorial on policy and technologies of this paper have the potential to kickstart future research and regulations that promote safe coexistence between active and passive users above 100~GHz, further benefiting the development of digital technologies and scientific exploration.

\end{abstract}


\begin{picture}(0,0)(10,-585)
\put(0,0){
\put(0,0){\footnotesize This paper has been accepted for publication in the Proceedings of the IEEE. Copyright IEEE, 2023.}
\put(0,-10){\scriptsize Personal use of this material is permitted. Permission from IEEE must be obtained for all other uses, in any current or future media, including reprinting/republishing}
\put(0,-20){\scriptsize this material for advertising or promotional purposes, creating new collective works, for resale or redistribution to servers or lists, or reuse of any}
\put(0,-30){\scriptsize copyrighted component of this work in other works.}}
\end{picture}

\glsresetall

%
\IEEEpeerreviewmaketitle

\section{Introduction}
\label{sec:introduction}


The digital society of the next decade will increasingly rely on services provided by a fundamental, invisible, yet scarce resource, i.e., the electromagnetic spectrum. 
The wireless spectrum enables a diverse set of applications, from high-speed communications~\cite{akyildiz2014terahertz}, to imaging~\cite{jepsen2011terahertz}, remote sensing~\cite{dickie2011thz}, Earth and space exploration~\cite{waters2006mls}, and radio astronomy~\cite{taylor1999synthesis}. 
\gls{rf} technologies, either communications or sensing, generally benefit from using a larger bandwidth, with a proportional improvement to the capacity and/or the sensing resolution~\cite{shannon1948mathematical,lee2017polarimetric}.
The finite nature of the spectrum, however, translates into limited availability of resources for each use case, prompting researchers to expand into new, unexplored portions of the spectrum, and to investigate sharing mechanisms. 
Notably, the crowded spectrum in the traditional sub-6 gigahertz (GHz) band has led \gls{5g} networks as well as radars and sensing systems to use the lower \gls{mmwave} band. For example, \gls{3gpp} NR, a \gls{5g} technology, is expected to use carrier frequencies as high as 71~GHz~\cite{38808}, while radars generally operate in multi-GHz bands around 60~GHz and 77~GHz~\cite{hasch2012millimeter}. Similarly, policies and technologies have evolved to accommodate shared uses of the spectrum~\cite{voicu2019survey}.

As the design and engineering of antennas and \gls{rf} circuitry for high-frequency transceivers and sensors has improved, the spectrum above 100~GHz has entered the spotlight as an enabler of \gls{6g} communications technologies, on the one side~\cite{jornet2011thz}, and of more advanced sensing solutions, on the other~\cite{mittleman2013sensing}. This portion of the spectrum includes large chunks of unused bandwidth, which both sensing and communications can benefit from.

Notably, certain sensing techniques integrate a large number of observations over time and frequency. As per the central limit theorem, the accuracy of the measurement is proportional to the square root of the number of observations~\cite{taylor1999synthesis}. Given a certain noise level target, a larger bandwidth allows measurements to be collected in a shorter time period, which is particularly beneficial in case of dynamic sources or measurement setups. Similarly, radars can provide more precise tracking with a larger bandwidth~\cite{caris2014very}. Additionally, the spectrum above 100~GHz features specific sub-bands of interest to the scientific community, because of molecular transitions~\cite{kokkoniemi2016discussion,waters2006mls}. 
Finally, larger bandwidth enables higher data rates for communications, with tens of GHz needed to target the terabit-per-second (Tbps) data rate goal of future \gls{6g} networks~\cite{giordani20196g,polese2020toward}, as we discuss later in this paper. The need for ultra-high data rates is also shared by the scientific community, to overcome limitations introduced by low capacity telemetry links between sensing satellites in space and ground stations~\cite{shehab2020recurrent}.


\begin{figure*}[h!]
\centering
	\includegraphics[width=\linewidth]{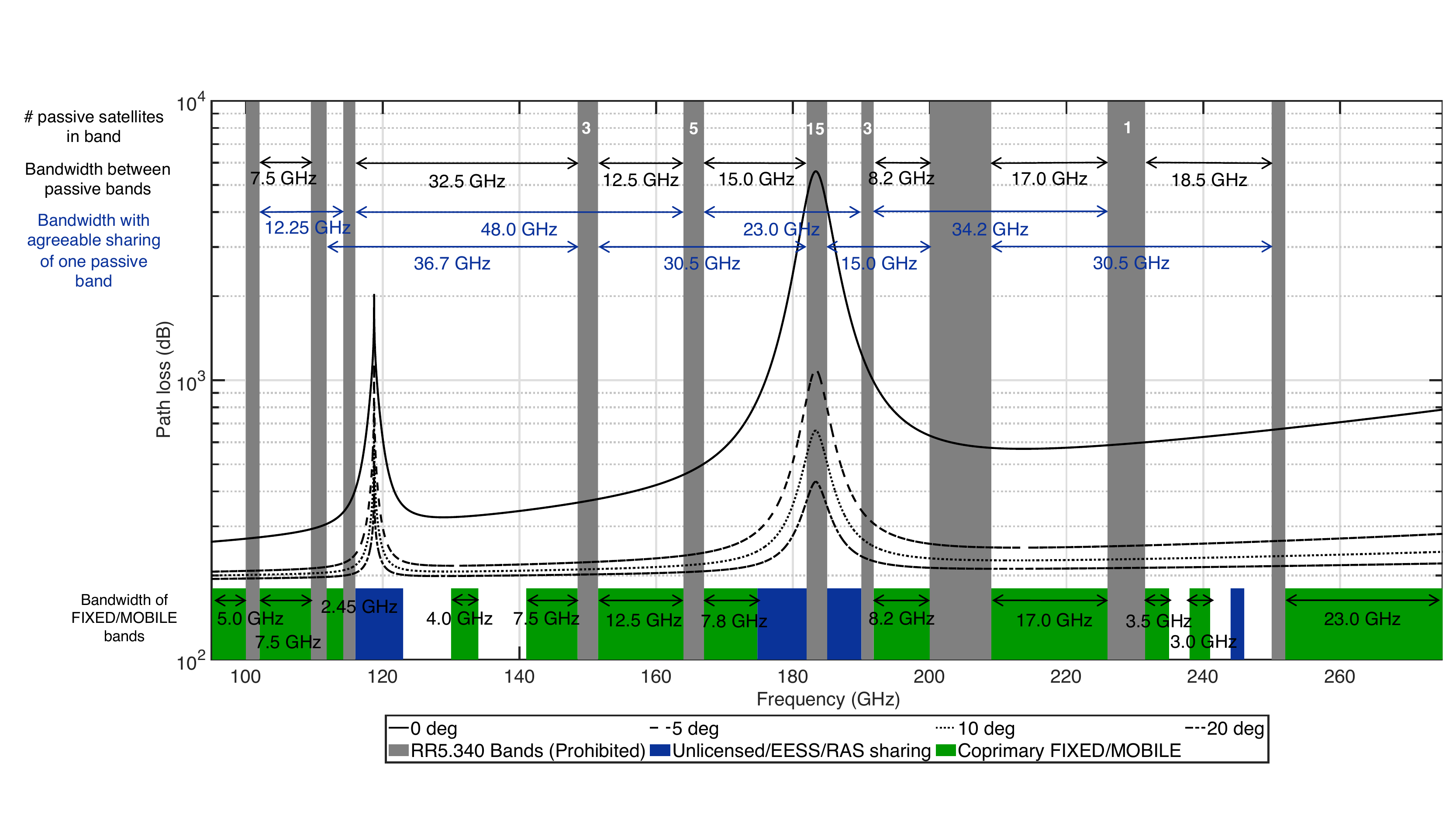}
	\caption{Spectrum allocation above 100~GHz and the path loss for a 407 km Earth-space link with different elevation angles\new{, based on the channel model discussed in Section~\ref{sec:interference}}. The figure also describes existing services \new{and spectrum allocation based on the \gls{itu} radio regulations~\cite{itu-radio-reg}}, and includes the number of passive satellites in each band dedicated to passive sensing. Finally, \new{the top part of the figure reports} the chunks of bandwidth that are available between passive bands (black) or by enabling sharing of one passive band (blue).}
	\label{fig:spectrum}
\end{figure*}

\subsection{Sharing the Spectrum in the 21st Century}

As a consequence, {\it both sensing and communications communities have a stake in the spectrum above 100~GHz}, either because of the availability of large chunks of untapped bandwidth, or because of specific frequencies of interest. The different communities, however, often express different needs with respect to how the spectrum should be used. For example, even modest levels of \gls{rfi} can strongly affect the performance of most classes of passive sensing systems \cite{marcus2014harmful}, while communications or other sensing techniques (e.g., some radars) need active transmissions. This has led national and international spectrum regulation entities to define a rigid allocation scheme in the spectrum above 100~GHz, with a set of narrow, yet numerous sub-bands exclusively dedicated to passive sensing (i.e., for science and space exploration)~\cite{itu-radio-reg}. While the total bandwidth reserved for passive users is rather small (i.e., 33.35 GHz between 100 and 275 GHz), the positioning of these sub-bands in the overall spectrum prevents the allocation of large chunks of contiguous bandwidth for active usage. \new{As highlighted in Figure~\ref{fig:spectrum}, the largest contiguous chunk allocated to active users between 100 and 275 GHz has a bandwidth of 32.5 GHz (between 116 GHz and 148.5 GHz). However, in this band, only 12.25 GHz (non contiguous) are allocated to fixed and mobile communications (i.e., 122.25-123 GHz, 130-134 GHz, 141-148.5 GHz), and the remaining 19.5 GHz are earmarked for active sensing techniques, radionavigation, and radiolocation.} For communications, this is less than the bandwidth that is typically needed to achieve the 1~Tbps target rate of future 6G networks. Similarly, the needs of the sensing community have evolved since when the current regulatory approach was established. At that time, the electronics for sensing systems were severely limited in terms of the bandwidth they could handle, and millimeter band astronomy facilities were not sensitive enough to see molecular features at large distances, where the expansion of the Universe changes their wavelengths substantially.
All of these aspects of the situation have now changed, so that the current approach of protecting narrow bands everywhere all the time is no longer as valuable as protecting broader bands at specific places and times.


Therefore, a dynamic and shared use of the spectrum above 100~GHz will simultaneously  enable next-generation sensing {\it and} communication technologies. Spectrum sharing has proven a successful enabler of advanced spectrum usage in the sub-6 GHz and lower \glspl{mmwave} bands. A notable example is that of the \gls{cbrs} band, i.e., 150 MHz between 3.55 GHz to 3.7 GHz which have been re-purposed for shared usage between sensing systems (e.g., U.S. Navy radars), critical communications (to and from aircrafts), and commercial users~\cite{sohul2015spectrum,fcc2015gn12354}. According to the recent regulations for \gls{cbrs}, secondary users (generally, commercial ones) need to preempt spectrum usage and effectively avoid \gls{rfi} to the primary users. This shows that it is possible to go beyond a strict resource partitioning scheme with exclusive spectrum access, with a mutually beneficial outcome, i.e., more bandwidth can be {\it dynamically} allocated to different services according to need. Along this line, a National Research Council report~\cite{nrc2010spectrum} shows that---with proper coordination---a constellation of 30 sensing satellites and active users on the ground can share the same bands with active transmissions for 99.7\% of the time and without any interference to the passive sensing system. The \gls{itu} has also advanced a resolution at the \gls{wrc} in 2000 to promote sharing of the passive bands above 71 GHz, as long as sharing does not expose passive services to interference levels beyond those standardized by \gls{itu}~\cite{itu2012perf,itu2003protection,itu2015levels,itu2019res731}.

The characteristics of devices and propagation in the above 100~GHz introduce new opportunities and challenges with respect to spectrum sharing approaches in the sub-6 GHz band~\cite{akyildiz2022terahertz}, \new{considering approaches that are based on device design, signal processing at the physical layer, and coordination at the \gls{mac} and above. The higher spreading and absorption losses} above 100~GHz~\cite{jornet2011thz}, together with directional transceiver architectures~\cite{andrello2018dynamic}, allow for increased spatial reuse and coexistence.
Similarly, a larger bandwidth can shorten the observation time for sensing and the transmission time for communications, thus potentially enabling more refined and flexible time-sharing strategies. 
At the same time, however, the device and RF circuit design is made more challenging by the high carrier frequency and large bandwidth processing requirements. \new{This complicates the design of precise transmit and receive frequency masks}, leading to out-of-band emission issues. \new{In addition, the extreme directionality required to close the link in this portion of the spectrum introduces a deafness problem which complicates the design of sensing- and beaconing-based sharing even when compared to \gls{mmwave} frequencies}.

\subsection{Contributions and Paper Structure}

The implementation of spectrum sharing techniques, therefore, requires a concerted effort involving the different stakeholders of the spectrum above 100~GHz, and innovations in both spectrum policy and engineering. This paper represents a critical step to make this portion of spectrum ready for the technologies that will use it in the future, replacing a set of regulations that date back to the 1930s~\cite{communications-act-1934,marcus2012spectrum}.

This paper provides policy and technological guidance on how sharing can be effectively implemented for the benefit of all the stakeholders in the spectrum above 100~GHz, enabling the digital revolution of the next decade. Notably, it is the first paper that:
\begin{enumerate}
    \item describes the needs of the stakeholders in the spectrum above 100~GHz, detailing---with numerical examples---why both sensing and communications can benefit from access to large, contiguous chunks of bandwidth in this portion of the spectrum;
    \item provides a detailed description of the current spectrum regulations in this band, highlighting possible policy roadblocks that prevent dynamic spectrum sharing between different stakeholders;
    \item adopts a physics-based approach to model the interference between active systems and passive users, to understand which scenarios and operating regimes are subject to \gls{rfi} in the spectrum above 100~GHz. The modeling will consider realistic settings and conditions, i.e., high-sensitivity receivers for the sensing systems, and directional antenna arrays in the communication systems, and will be based on \gls{itu} channel models and antenna patterns. Our analysis highlights that while the high path loss and directional antennas can help reduce RFI in some scenarios, there are configurations (e.g., for high satellite elevation angles) where \gls{rfi} from terrestrial communications links exceeds the harmful \gls{rfi} thresholds for passive services set by the ITU;
    \item analyzes technological enablers of spectrum sharing in the spectrum above 100~GHz, \newrev{considering established techniques (e.g., signal processing for \gls{rfi} mitigation), specific solutions at higher frequencies (e.g., metasurfaces and extremely directional arrays), and research directions and challenges toward spectrum sharing above 100 GHz. We review}
    full-stack solutions, i.e., hardware-based \gls{rfi} mitigation (innovative antenna design, antenna arrays, \glspl{fss}), signal processing, and communications and networking design;
    \item discusses how spectrum regulations and technologies can evolve to accommodate more shared spectrum, proposing a future directions for technology and policy development, experimentation, and commercialization of sensing and communications solutions above 100~GHz.
\end{enumerate}

\begin{figure}[t!]
	\centering
	\includegraphics[width=.85\columnwidth]{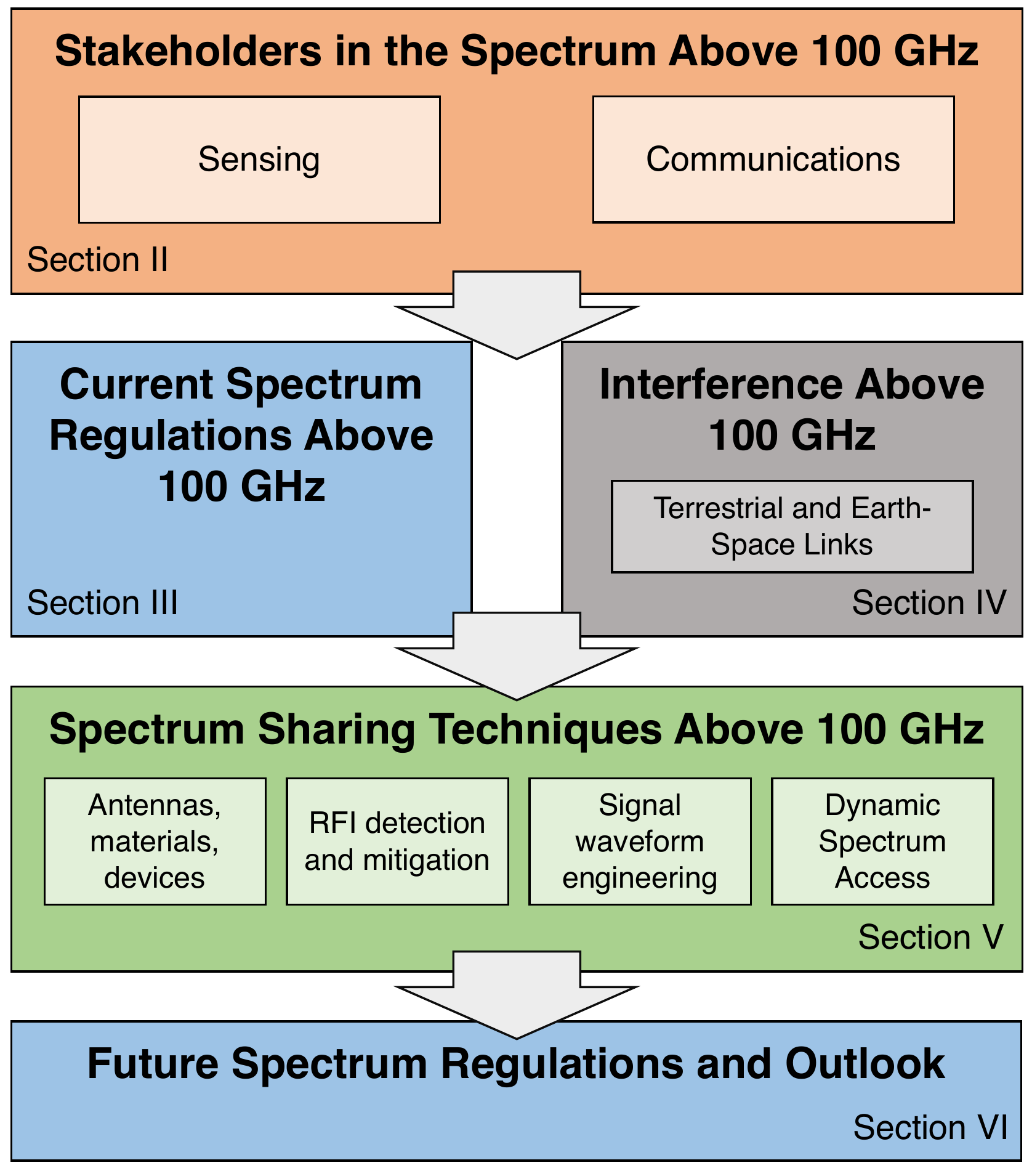}
	\caption{Structure of the paper}
	\label{fig:paper_structure}
\end{figure}

The remainder of this paper is organized as follows (see Figure~\ref{fig:paper_structure}). Section~\ref{sec:stakeholders} describes why the spectrum above 100~GHz is of interest for both the sensing and communication communities, and why more bandwidth is generally useful for both. Section~\ref{sec:current_regulations} outlines the current regulations that prevent a flexible use of the spectrum above 100~GHz, while Section~\ref{sec:interference} presents numerical results on the interference that can arise in different scenarios. Based on this analysis, Section~\ref{sec:sharing} introduces and reviews possible spectrum sharing techniques \new{and related research challenges}. Finally, Section~\ref{sec:conclusion} concludes the paper with a discussion on what is required to enable spectrum sharing above 100~GHz.

\section{Stakeholders in the Spectrum Above 100~GHz}
\label{sec:stakeholders}

\begin{figure}[t]
	\centering
	\includegraphics[width=\columnwidth]{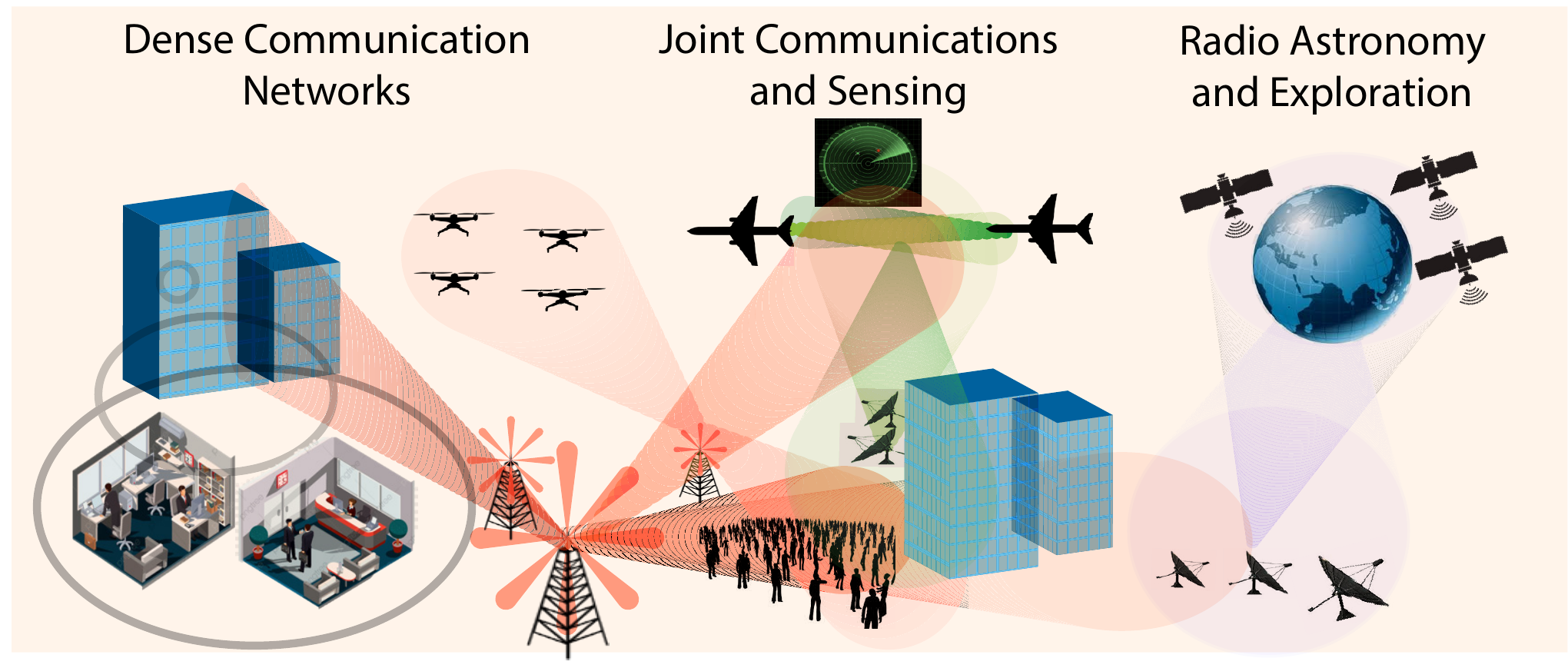}
	\caption{Stakeholders in the above 100~GHz band.}
\label{fig:stakeholders}
\end{figure}

As shown in Figure~\ref{fig:stakeholders}, the spectrum above 100~GHz is an attractive resources for a diverse set of stakeholders. In the following paragraphs, we will discuss why each user of the spectrum is interested in using a large bandwidth, \new{focusing specifically on the spectrum above 100 GHz and providing a quantitative assessment when relevant}.

\subsection{Sensing}
\label{subsec:sensing}
Sensing applications which require \new{access to spectrum} consist of three broad categories: radar systems, passive Earth-observing systems, and astronomy systems. Broadband (or continuum) and narrow-band applications are both important in the Earth-observing and astronomical contexts. 

The narrow-band work focuses on detection of specific molecules.  Key molecules for atmospheric science include molecular oxygen (at 60 GHz) and water vapor (at 183 GHz). For astronomy, the most important molecule has traditionally been carbon monoxide (CO) (at 115, 230 and 346 GHz), which is the most abundant asymmetric molecule, and hence the strongest emission feature from regions with dense gas. 

Broad bandwidths can be desirable for molecular line astronomy. This is because of Doppler motions and general relativistic effects, due to the expansion of the Universe, which can have a significant impact by moving the sensing target out of typical narrow bands. With facilities that exist or are planned for the near future, the strongest molecular features may be detectable up to distances sufficiently close to the edge of the observable Universe. In this case, their wavelengths may be a factor of about 4 larger as we observe them than the wavelengths of the same molecular transitions as measured in a laboratory. This makes the case for a broadband access to the spectrum for radio astronomy sensing. Additionally, the increasing sensing capabilities of modern scientific facilities have allowed observers in recent years to measure phenomena related to a much larger number of rare molecules, which span the whole electromagnetic spectrum. In general, projects which involve characterization of the emission of a single known molecule can usually do well with narrow-band observations, while projects which involve searches for new molecules typically need broadband access to the spectrum, even if the features to be observed are narrow ones. 

Typical spectral widths of the features vary by application.  Atmospheric features are susceptible to pressure broadening (i.e., the intrinsic energies of the transitions in one molecule can be affected by forces exerted from other molecules), and sometimes Zeeman broadening (i.e., they are broadened by forces from the Earth's magnetic field affecting the energy levels within molecules).  Additionally, some atmospheric molecular features come in bands with many adjacent transitions.  The oxygen absorption band, for example, consists of a large set of absorption features, pressure broadened to create a smooth feature from about 50-70 GHz, and the water vapor band at 183 GHz also spans a fairly broad range, dominating atmospheric opacity from about 165 GHz to 200 GHz~\cite{prigent2006comparisons}.  Astronomical narrow-band features typically have widths no more than those of the Doppler shifts associated with the orbital speeds of galaxies.  These are typically $\approx$~200-400~km/sec, giving fractional widths of about 1/1000, or a few hundred MHz.  There are, however, also often isotopic features adjacent to the primary molecular feature that astronomers wish to observe simultaneously with the primary feature. These have fractional separations about the same as the fractional mass difference between the dominant and rarer isotopes, so they are typically shifted by about 5\% from the main feature for the light molecules that are most abundant (e.g., in addition to CO at 230 GHz, there are $^{13}$CO at 220 GHz and C$^{18}$O at 210 GHz).

Finally, broadband passive work also focuses on detection of thermal and synchrotron emission.\footnote{Synchrotron emission is the relativistic version of cyclotron emission, and is produced when electrons with speeds very close to the speed of light produce radiation \new{as} they are accelerated by magnetic fields.}  Broadband active sensing work largely involves the use of the millimeter band to do the same kind of work as lower frequency radars, taking advantage of the short wavelength at these frequencies to perform finer imaging.  Broadband passive systems generally study phenomena with emission across the entire millimeter band, so the restrictions of sensing applications arise from the limitations on the receivers and the electronics and computational processing systems.  Currently, these generally impose 20\% fractional bandwidths on the receivers, and 8-50 GHz total bandwidth from the electronics. Nonetheless, technological improvements have consistently pushed these limits over the past few decades, and there are no indications that such progress is slowing down. The spectrum above 100~GHz is also used for some very short range imaging applications (e.g. medical imaging and security scanners).

In the following paragraphs, we will discuss in details use cases for narrow-band and continuum sensing.

\textbf{Narrow-band applications.}
The narrow-band applications are largely driven by a particular set  of frequencies for specific rotational transitions of molecules, largely within the 10-500~GHz range. Earth-sensing satellites and high altitude airplanes can use these molecular features to map out quantities such as chemical composition of the atmosphere and cloud cover and humidity, among others~\cite{VIPR,skofronick2017global,tobin2006atmospheric}. Such measurements are important for both meteorology and climate science.  

For astronomical objects, molecular features can provide information not just about chemical compositions, but also temperatures and densities of regions in space.  Temperatures are largely probed by the fraction of a single compound's molecules in different excited states, while densities and chemical compositions can then be inferred from the relative abundances of different molecules.  In recent years, improvements in the sensitivity of astronomical facilities in the 10-500~GHz band have made it possible for astronomers to detect a wide range of new molecules~\cite{mcguire_chemicals}. The newest facilities can also detect molecules with the brightest features at distances where the expansion of the Universe introduces large cosmological redshifts~\cite{combes_review}; these large cosmological redshifts can also move important molecular transitions out of protected bands, into unallocated frequency ranges.  Both of these developments mean that even the portion of the passive sensing community working with narrow spectral features has \textit{applications that span the full spectrum}.

\textbf{Broadband (or continuum) sensing.}
There is considerable interest in moving automotive and other radar systems above 100~GHz, although this work is still in relatively early stages of development~\cite{autoradar,military_radar}\new{, as the larger bandwidth and shorter wavelength lead to better range, angular and velocity resolutions.} The range resolution \new{$\Delta R$} of a radar based on pulse compression depends on the bandwidth of the transmitted pulse, i.e., 
\begin{equation}
    \Delta R = \frac{c}{2B},
\end{equation}
where $c$ is the speed of light and $B$ is the bandwidth. The angular resolution depends on the antenna beamwidth, which is related to the electrical length and corresponding directivity of the antenna. The velocity resolution depends on the Doppler spectrum that is also related to the wavelength of the transmitted signal.
Moreover, in some cases, the propagation challenges of working in the above 100~GHz band are a positive feature; a higher density of automotive radar systems can be tolerated if propagation losses keep the systems from interfering with one another over large distances.

Broadly speaking, however, the continuum work presently done in this band is radio astronomy.  Continuum work in radio astronomy is usually done to look at cool thermal emitters, such as (i) interstellar dust; (ii) the disks around young stellar objects, in which planets form (which typically have temperatures of a few hundred Kelvin); or (iii) the cosmic microwave background (which has a temperature of 2.7 K). It can also be used to look at sources of synchrotron emission at high frequencies. The highest angular resolution images that can be made by astronomers come from millimeter-band very long baseline interferometry, work done at 230 GHz~\cite{EHT}. At longer wavelengths, the angular resolution is insufficient to measure black hole shadows without space-based antennas (which currently lack the sensitivity to do so).  At infrared bands, interferometry has only been possible with direct interference of signals from physically connected telescopes, limiting baselines to a few hundred meters.    The millimeter band is the shortest wavelength on which data can be recorded and global baselines can be achieved. This capability allowed---for example---the measurement of the size of the shadow of a black hole event horizon in the galaxy M87 \cite{EHT}.  In most cases of continuum imaging, the largest possible bandwidth is desired in order to detect faint signals, but for sufficiently bright sources, with sufficiently large bandwidths, the spectral shape of a source can be measured with a single observation at a single band.

The sensitivity of a radiometer is derived from the radiometer equation:
\begin{equation}
    \Delta T = \frac{T_\text{sys}}{\sqrt{B \tau}},
\end{equation}
where $\Delta T$ is the residual (root-mean-square) uncertainty in a noise temperature measurement, $T_\text{sys}$ is the noise temperature of the system being measured, and $B \tau$ is the product between bandwidth $B$ and time $\tau$ of observation~\cite{hunter2015statistical}.  The signal-to-noise of a source is then derived from the ratio of the source brightness temperature, $T_\text{source}$, and $\Delta T$. Notably, $T_\text{source}=\left(\frac{A_\text{e}}{2k_\text{B}}\right)S_\nu$, where $A_\text{e}$ is the effective area of the system, $k_\text{B}$ is the Boltzmann constant and $S_\nu$ is the flux density of the source per angular resolution element.

Radio astronomy requires ultra-low noise levels.  Reaching these in short exposure times is always desirable, as, at a minimum, it enables a larger number of  observations to be done. Additionally, in some cases, strong source variability is the goal of the measurement itself~\cite{tetarenko}, with the consequence that the measurement must be completed quickly in order to make reliable aperture synthesis images~\cite{EHT}.  Increases in sensitivity in a given exposure time can be achieved either by increasing the bandwidth available or by increasing the effective area of the system.  Bandwidth increases are generally preferred because electronics upgrades are usually much less expensive than adding antennas or increasing the sizes of the antennas used.  Available bandwidth currently varies among the different telescopes.  The receiver bands are generally limited to about 20\% of the central frequency.   At the present time, millimeter band interferometers  typically use about 8~GHz, with the specific frequencies chosen based on the science case for the user.\footnote{https://almascience.eso.org/about-alma/alma-basics}\textsuperscript{,}\footnote{https://www.iram.fr/IRAMFR/GILDAS/doc/pdf/noema-intro.pdf} 
However, for the Atacama Large Millimeter Array, planning is already underway to upgrade the correlators to allow at least a factor of 2 increase in bandwidth~\cite{ALMA_roadmap}.  Single dish radio telescopes have larger bandwidths; for example, the TolTEC instrument for the Large Millimeter Telescope planned to reach about 60~GHz of bandwidth.\footnote{https://people.astro.umass.edu/~wilson/TolTEC\_website/about.html}
The present regulatory framework was established in an era long before the current tens of GHz of bandwidth systems were practical.

Currently, a rather limited number of radio astronomy facilities engage in sensing above 100~GHz.  All such facilities are in high altitude, dry locations, generally several kilometers or more from settled locations; the requirement of detecting faint signals mandates choosing sites with minimal propagation losses.  A substantially larger (albeit still small) number of telescopes work in the 90 GHz band, and the CLOUDSAT satellites of NASA have successfully avoided interfering with them for 15 years.  It should be straightforward to continue to avoid interference with ground-based scientific users.

\subsection{Communications}

Several early papers on \gls{6g} communication technologies have highlighted how the spectrum above 100~GHz could be a fundamental enabler of ultra-high-capacity wireless networks~\cite{polese2020toward,akyildiz2022terahertz,rappaport2019wireless}. Results and demonstration of the feasibility of communicating in the (sub) terahertz bands have also been shown in several works, for example as in~\cite{castro2020experimental}. The main challenges for communications in this portion of the spectrum are associated to (i) the sensitivity to blockage and the high propagation loss~\cite{jornet2011thz}, which can be offset, however, using directional antennas; (ii) the presence of frequency selective fading in specific bands, which increases with the humidity; and (iii) the engineering and manufacturing of embedded \gls{rf} circuitry that can operate at such high frequencies~\cite{pfeiffer2018current}. Additional challenges are related to the interfacing with the higher layers of communications protocol stacks, such neighbor and infrastructure discovery challenged by directional links, and processing data at ultra-high rates~\cite{polese2020toward}.

Nonetheless, the need for Tbps data rates in 6G networks makes a compelling use for the deployment of communication services in the spectrum above 100~GHz, with a vast available bandwidth. Indeed, \new{the achievable data rates $R$ of a wireless link (with an ideal, frequency-flat channel)} is proportional to the bandwidth $B$, \new{scaled by} the spatial reuse or diversity factor $K$ and the spectral efficiency $S$, and \new{by a function $f(\cdot)$ of} the link \gls{sinr} $\Gamma$:
\begin{equation}\label{eq:shannon}
    \new{R} \propto K \cdot S \cdot B \cdot \new{f(\Gamma).}
\end{equation}
Here, $f(\cdot)$ is a generic function expressing the relationship between \gls{sinr} and achievable rate, \newrev{and the different elements $K, S, B$} in Eq.~\eqref{eq:shannon} highlight how next-generation communication technologies can improve on existing ones to increase the \new{communication rates}. Notably, a higher spectral efficiency can be achieved through more spectrally-efficient modulations, e.g., \gls{ofdm} and its variants~\cite{benvenuto2021algorithms,banelli2014modulation,berardinelli2014potential}, combined with high order \gls{qam} (with 1024-QAM and even above). \newrev{This is represented in Eq.~\eqref{eq:shannon} by the scaling factor $S = \log_2(M)$, with $M$ the modulation order.} Spatial reuse and diversity can be achieved using \gls{mimo} techniques, which allow multiple data streams to be transmitted over the same time and frequency resources. This is represented in Eq.~\eqref{eq:shannon} by the scaling factor $K$, which we model as the number of MIMO streams, proportional to the rank of the channel matrix. A recent \gls{mimo} evolution, i.e., massive \gls{mimo}~\cite{marzetta2011noncooperative}, has been shown to improve the \gls{sinr} term as well, by exploiting the much larger number of antennas at the transmitter as compared to the receiver. The idea of \gls{ummimo}~\cite{akyildiz2016realizing} extends this even further, by packing thousands of antenna elements in small, portable arrays. Small cells and ultra-dense networks also help increase the spatial reuse factor $K$, sometimes at the cost of increased interference which penalizes the term $\Gamma$. Similarly, increasing the communication bandwidth $B$ 
allows for a linear increase in the product outside function $f(\cdot)$, but at the same time, the noise term of the \gls{sinr} $\Gamma$ increases proportionally to $B$.

\begin{figure}[t]
\ifexttikz
    \tikzsetnextfilename{datarates-mmimo}
\fi
\begin{subfigure}[t]{\columnwidth}
\centering
  \setlength\fwidth{.95\columnwidth}
  \setlength\fheight{.6\columnwidth}
  \input{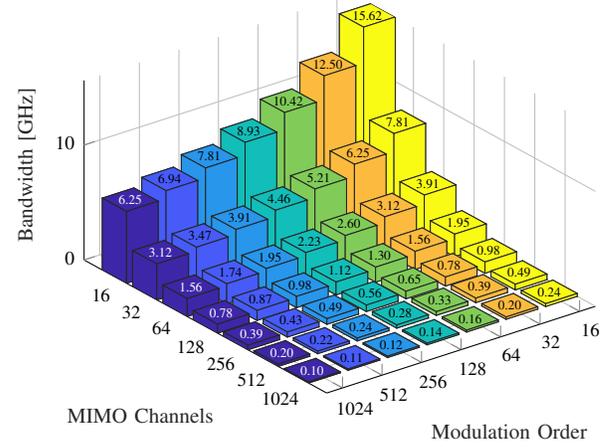}  
	\caption{High number of \gls{mimo} streams (e.g., enabled by massive \gls{mimo}).} 
	\label{fig:datarates_mmimo}
\end{subfigure}
\ifexttikz
    \tikzsetnextfilename{datarates-mimo}
\fi
\begin{subfigure}[t]{\columnwidth}
\centering
  \setlength\fwidth{.8\columnwidth}
  \setlength\fheight{.6\columnwidth}
  \input{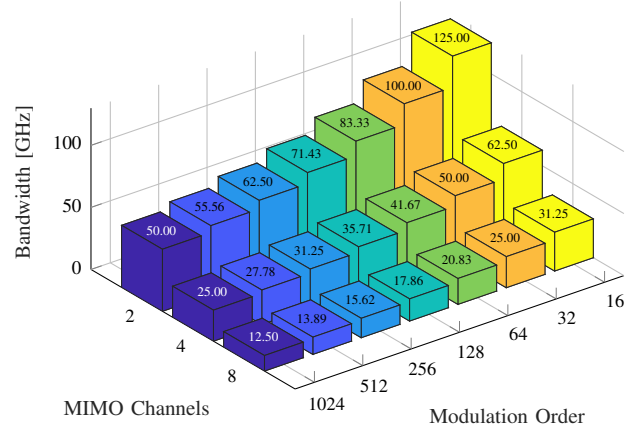}	
  \caption{Low number of \gls{mimo} streams.}
	\label{fig:datarates_mimo}
\end{subfigure}
\caption{Required bandwidth to achieve a 1 Tbps data rate, as a function of modulation order and MIMO channels for OFDM.}
\label{fig:datarates}
\end{figure}

\begin{table*}[t]
    \centering
    \begin{tabular}{lllllll}
        \toprule
         Band (GHz) & 122.25-130 & 158.5-164 & 167-174.8 & 191.8-200 & 231.5-235 & 238-241  \\\midrule
         Bandwidth (GHz) & 7.5 & 5.5 & 7.8 & 8.2 & 3.5 & 3  \\
         Primary allocation & Fixed-Satellite/Mobile-Satellite & Fixed/Mobile & Fixed/Mobile & Fixed/Mobile & Fixed/Mobile & Fixed/Mobile \\
         \bottomrule
    \end{tabular}
    \caption{Bands above 100~GHz with Mobile or Fixed allocations without passive allocations. Most of these bands have co-primary allocations with other active services.}
    \label{table:allocations}
\end{table*}


The data rate increase for 6G networks will be given by a combination of the elements above. Nonetheless, the bandwidth plays a key role, as we show in Figure~\ref{fig:datarates}. Each bar in the the figure represents the bandwidth required to achieve a data rate of $R_t = 1$~Tbps, which is expected to be one of the defining \glspl{kpi} of 6G networks~\cite{giordani20196g,rajatheva2020scoring} \new{for a number of applications, including backhaul~\cite{palicotdeliverable,rajatheva2020scoring,kurner2022demonstrating}, immersive, holograph-based virtual experience~\cite{xu20113d}, data offloading in vehicular networks~\cite{samy2021power}, and mobile hotspot scenarios~\cite{rajatheva2020scoring}.} \new{Notably, the bandwidth is computed as the ratio between the target rate $R_t$ and the efficiency $\eta$ of the MIMO and modulation order in use, i.e., $\eta = \hat{K} \hat{S} / \alpha$, with $\hat{K}$ the number of \gls{mimo} streams, $\hat{S}$ the bit per symbol rate of a certain modulation order, and $\alpha=1$ for single-sided waveforms, 2 otherwise.}
Figure~\ref{fig:datarates_mmimo} considers a configuration with massive \gls{mimo} capabilities, and highlights how an extremely high number of \gls{mimo} channels, combined with a high modulation order, makes it possible---in principle---to reach the data rate target with bandwidths that are comparable to those of current 5G systems. Designing and engineering a system with such capabilities, however, is not trivial. A large number of \gls{mimo} streams requires a massive number of antennas, as well as diversity in the channel conditions, and this may not be always feasible, especially when considering point-to-point backhaul links or portable devices in the sub-6 GHz spectrum.

Portable arrays with hundreds of antenna elements are instead feasible above 100~GHz, thanks to the small wavelength which translates into small antenna elements~\cite{akyildiz2016realizing}. At such high frequencies, however, it becomes extremely complex to design multi-channel and/or multi-user \gls{mimo} systems with a high number of channels. The state of the art of the research on baseband design for such systems is represented by 4x4 and 8x8 \gls{mimo} systems~\cite{simsek2018140,kueppers2017compact}. To this end, Figure~\ref{fig:datarates_mimo} considers configuration with a more practical number of parallel \gls{mimo} streams (i.e., up to 8). In this case, even with an OFDM modulation with 1024 QAM per channel, more than 12.5 GHz of contiguous bandwidth are needed. 

Additionally, such constellations often require extremely good channel conditions and low noise (both in amplitude and phase). Therefore, more practical configurations are represent on the right part of Figure~\ref{fig:datarates_mimo}, with 16 QAM, which further increases the requirement in terms of bandwidth (i.e., 31.25 GHz with 16 QAM and 8x8 \gls{mimo}). Therefore,
{\it the need for ultra-wide bandwidth makes the spectrum above 100~GHz a key enabler for future 6G communications.}

\section{Current Spectrum Regulations above 100~GHz}
\label{sec:current_regulations}

Enabling sharing of the spectrum above 100~GHz will require both technological progress and policy updates. This section reviews the development of spectrum policy for the bands between 100~GHz and 275~GHz (as there are no formal \gls{itu} allocations above 275~GHz yet), highlighting the roadblocks and opportunities towards a sharing-conscious spectrum policy. \new{Notably, we review and summarize international regulations and provide an historical perspective on how they have been developed, together with a discussion on how the characteristics of the spectrum above 100 GHz have led to different policies when compared to sub-6 GHz or lower \gls{mmwave} spectrum.}

\subsection{International Telecommunications Union (ITU)}

At the international level, the spectrum is regulated by the \gls{itu}, a United Nations agency with 193 Member states that develops the ITU \gls{rre}, a treaty document generally binding on its signatories.  While these rules do not require countries to use spectrum in certain ways, they prohibit them from causing ``harmful interference'' to spectrum uses of other countries which are fully consistent with the rules.  While most regulations involve active transmitters for applications such as telecommunications, broadcasting and radar/radiolocation, the ITU also recognizes and protects passive scientific uses of the spectrum. As discussed in Section~\ref{sec:stakeholders}, these use cases involve only the reception of naturally occurring emissions in specific spectrum bands, for example for radio astronomy and environmental monitoring.
Notably, the spectrum above 100~GHz features several bands which are reserved for the passive community. Table~\ref{table:allocations} and Figure~\ref{fig:spectrum} list the bands with primary or co-primary Fixed and/or Mobile allocations, without any co-primary passive allocation:\footnote{\new{Primary users are the only entities that can legitimately use a portion of the spectrum and are entitled to protection from harmful \gls{rfi} from secondary users. In case of multiple co-primary allocations, the co-primary users need to coordinate with each other to avoid harmful \gls{rfi}.}} in the present allocation scheme, the maximum fixed or mobile congruous bandwidth below 252 GHz is 8.2 GHz. This, as previously discussed, motivates the development of spectrum sharing solutions.

\subsection{WRC-2000 and spectrum policy above 100~GHz}

Prior to the ITU \gls{wrc} in 2000, ITU had few rules and allocations dealing with spectrum above 100~GHz.  At that conference, similar proposals from both the United States and the \gls{cept} advocated the basic policies now in place in 100-275~GHz~\cite{itu2000wrc}.  The policies advocated by the US and Europe included both numerous passive allocations above 100~GHz, some bands with no passive allocations at all, and some bands in which active services and passive services could nominally share on a ``co-primary'' basis. 

Ten passive-only bands were created in the 100-275 GHz spectrum, using the restriction in footnote 5.340 of the Allocation Table (also known as \gls{rre}5.340)~\cite{itu-radio-reg}, previously used for lower frequency passive bands. \gls{rre}5.340 states that ``all emissions are prohibited'' in these bands, which cover 19\% of the 100-275 GHz spectrum. Table~\ref{table:passive-allocations} compares the amount of ``prohibited'' spectrum in the \gls{ehf} band (30-300 GHz) with lower spectrum regions. While the amount of prohibited spectrum in \gls{uhf} and \gls{shf} (below 30 GHz) is negligible, it covers 15\% of the entire \gls{ehf} region.  In addition, the 15 protected bands in the \gls{ehf} spectrum fragment the band allocations in ways unknown in lower regions.  Finally, these passive allocations are more dense in the 100~GHz region, because molecular resonances that are of interest to astronomers and environmental scientists are more densely concentrated above 100~GHz than below~\cite{nrc2010spectrum}.

\begin{table}[t]
    \centering
    \begin{tabular}{llll}
        \toprule
          Band   &	Frequency (GHz) &	\begin{tabular}[x]{@{}l@{}}Number of RR5.340\\Passive Blocks\end{tabular}	& \begin{tabular}[x]{@{}l@{}}Fraction of\\Band Passive\end{tabular} \\ \midrule
         \gls{uhf} &	0.3$-$3 &	2 &	1\%  \\
         \gls{shf} &	3$-$30 &	3	& 2\% \\
         \gls{ehf} &	30$-$300 &	15	& 15\% \\\bottomrule

    \end{tabular}
    \caption{Prohibited bands according to \gls{rre}5.340 in different spectrum regions.}
    \label{table:passive-allocations}
\end{table}

\begin{table*}[t]
    \centering
    \glsunset{ras}
    \glsunset{eess}
    \begin{tabular}{lllllllll}
        \toprule
         Band (GHz) & 102-109.5 & 111.8-114.5 & 141-148.5.8 & 151-155.5 & 155.5-158.5 & 209-226 & 252-265 \\\midrule
         Bandwidth (GHz) & 7.5 & 2.45 & 7.5 & 4.5 & 3.5 & 3 & 13 \\
         \new{Active co-primary allocation} & \multicolumn{7}{c}{Fixed/Mobile} \\
         \new{Passive co-primary allocation} & \gls{ras} & \gls{ras} & \gls{ras} & \gls{ras} & \gls{ras}, \gls{eess} & \gls{ras} & \gls{ras} \\
         \bottomrule
    \end{tabular}
    \caption{Bands above 100~GHz with Mobile or Fixed allocations with \gls{ras} co-primary passive allocations.}
    \label{table:allocations-ras}
    
\glsreset{ras}
\glsreset{eess}
\end{table*}

WRC-2000 and succeeding conferences have defined 18 bands with co-primary passive allocations above 100~GHz. In some, passive allocations are co-primary with terrestrial fixed and mobile services, and in some others, only with other services, including inter-satellite links. These co-primary bands cover 58\% of the spectrum in this region.  As a result, 77\% of the spectrum in 100-275 GHz has either primary of co-primary passive allocations.



\gls{itu} regulations define 3 basic types of passive services with such allocations, that broadly classify the spectrum stakeholders described in Section~\ref{sec:stakeholders}: \gls{ras}, \gls{eess}, and \gls{ses}.  \gls{ras} is the oldest, and involves Earth-based sensors looking radio sources from sources far away from the Earth.  In practice at frequencies above 100~GHz atmospheric absorption of such signals is a major issue so the best locations are high altitude arid sites.  In particular, Chile’s Atacama Desert is a major location for such observatories.  There are few observatories in the spectrum in US territory and the only one near a US urbanized area is the Arizona Radio Observatory at Kitt Peak about 80 km from Tucson AZ. 
Traditionally, RAS observatories have been protected from active users of the same bands by coordinating the siting and parameters of transmitter within an established ``coordination zone''~\cite{itu200710312}. National spectrum regulators establish zones around radio telescopes in which all transmitters need prior discussion with the radio telescope operators before they can be installed. 
The size of these coordination zones depends on both the frequencies being used at the radio telescope and the specific terrain around it. 
\new{Indeed, \gls{ras} antennas point upward and thus can only receive power from terrestrial transmitters that are in or near a line of sight from the antenna, therefore assessing transmitter locations, direction, and power levels in coordination areas is of utmost importance}.
Table~\ref{table:allocations-ras} summarizes the allocations with \gls{ras} as \new{the co-primary user for Fixed/Mobile users}, which already potentially represent a first option for spectrum sharing with terrestrial active users. \new{Finally, an emerging source of possible \gls{rfi} for \gls{ras} antennas is represented by aerial vehicles, e.g., airplanes or \glspl{uav}, that maintain connectivity with the ground.}

\new{Differently from \gls{ras} systems,} EESS antennas are on non-geostationary satellites at orbit heights of 400-800 km  and point towards a large area on the Earth, although some limb sensing antennas point towards the horizon. EESS satellite makes observations throughout their orbits and thus could be affected by sidelobe or scattered power from terrestrial radio systems in their bands and thus are the most difficult class of passive systems to protect, both from regulation and technological standpoints.

The unusual density of passive bands above 100~GHz fragments the available spectrum, forbidding or complicating the use of large contiguous bandwidths.  These condition currently hold in the 100-275 GHz band, as also shown in Figure~\ref{fig:spectrum}:
\begin{itemize}
    \item the largest bandwidth with a present fixed or mobile allocation but no co-primary passive allocation is 8.2 GHz (191.8-200 GHz);
    \item the largest contiguous bandwidth possible for active use (but not necessarily fixed or mobile allocations) without overlapping a RR5.340 band is 32.5 GHz (116-148.5 GHz); and 
    \item the largest bandwidth for active use without overlapping any band with a \gls{eess} allocation is 26.25 GHz (122.25-148.5 GHz).
\end{itemize}
Thus use of bandwidth greater than 26.25 GHz for active use will require sharing with a primary or co-primary passive service, motivating the development of technological solutions for active/passive spectrum sharing. To extend active uses to 32.5~GHz, active and passive users will need to share RR5.340 bands. Doing so will not only require technological innovations, but also the development of new policies. An apt starting point for such policies would be research findings which indicate bands less likely to be subjected to interference, and be ratified by ITU-R. The process involves placing the issue in an agenda at a \gls{wrc} conference, followed by discussion and, (in case) approval at the next \gls{wrc} conference. It is worth noting that \gls{wrc} conferences happen every four years, which reduces the speed of the overall process.

\subsection{ITU-R Resolution 731 and Spectrum Sharing Policy}
\label{sec:burdgen}
At WRC-2000, the US and \gls{cept} proposals also included a provision to study whether sharing of the allocated passive bands with active services was possible without harmful interference.  In 2000, on the eve of the commercial introduction of 3G~/~\gls{umts} cellular technology, such studies were complicated by the fact that there were few indications of what meaningful telecommunications applications above 100~GHz, or even what data rates would be required in the long term, for either mobile links or the backhaul.
Resolution 731~\cite{itu2019res731} was approved as a compromise between the suggested language of the US and CEPT on spectrum sharing. It states that ITU-R should study possible sharing of the passive bands above 71~GHz with other spectrum uses, and that such sharing must not expose the passive service to interference levels greater than those specified by ITU-R recommendations~\cite{itu2012perf,itu2003protection,itu2015levels}. At WRC-19, Resolution 731 was then updated to add new provisions for above 275~GHz, but its provisions within the 71-275~GHz spectrum remained the same, except for updating references to other ITU-R documents. 

Resolution 731 specifies explicit interference criteria for \gls{eess}~\cite{itu2012perf} and for RAS~\cite{itu2003protection,itu2015levels}.  In the case of \gls{eess}, the criteria given is a triplet of numbers for each protected band which specifies: (i) the reference bandwidth over which interference power should be measured; (ii) the maximum permitted interference level; and (iii) the percentage of the Earth's surface area or time in which the permissible interference level may be exceeded.  These power levels are defined at the terminals of the receiving antenna in the satellite. They represent the total power received from the area of the Earth that is viewed by the antenna’s footprint, including its sidelobes.  A separate ITU-R document~\cite{itu2015typical} gives typical satellite parameters, such as orbit height and antenna characteristics but does not presently cover every possible band (although an update is in progress~\cite{itu2014rs2064}). The analysis of the available literature on \gls{eess}, however, does not exhaustively clarify if all passive satellites have interference vulnerability comparable to that specified in this document. It is also not clear whether there is public information available on all passive satellites entitled to interference protection and their parameters, including orbit parameters.

Finally, Resolution 731 also contains the following provision: ``to the extent practicable, the burden of sharing among active and passive services should be equitably distributed among the services to which allocations are made''~\cite{itu2019res731}.  This implies that, when developing sharing strategies, both the active and passive users have an obligation to cooperate to increase the overall use of the spectrum, while also achieving the user-specific objectives. 
Nonetheless, passive satellite systems do not have the same level of flexibility as active users. 
The design of satellites and scientific equipment is a multi-year effort, and careful planning needs to be factored in the engineering of the RF components. 
Once in orbit, there is little or no possibility to change operating parameters and adjust transceiver systems to react to new or unforeseen interfering signals. 
As of today, most passive satellites orbiting are expensive units, with a mission lifetime spanning 5 to 15 years. It is also difficult to change parameters for satellites that are planned but not launched yet, even if consensus on the policy and sharing technologies is reached. So far, there has been no ITU-R discussion on the concept of ``burden sharing'' as introduced in Resolution 731, even though this holds long term potential for improving spectrum sharing capability.

\section{Interference above 100~GHz: Link Budget Analysis}
\label{sec:interference}

As discussed in Section~\ref{sec:current_regulations}, there exist some bands in which the active users are forbidden from emitting any \gls{rf} signals. Herein, we present numerical results that highlight scenarios in which harmful interference arises, and deployments in which the same is rendered void. We first describe our modeling approach (Section~\ref{sec:system_model}), based on ITU channel models, and then the results (Sections~\ref{sec:pathloss} and \ref{sec:intf}). \new{Notably, we analyze terrestrial and aerial links as they represent typical communication and interference scenarios, e.g., with a backhaul link on the ground generating \gls{rfi} toward a satellite.}


\subsection{System Model}
\label{sec:system_model}

\begin{figure}[t]
    \centering
    \includegraphics[width=.9\columnwidth]{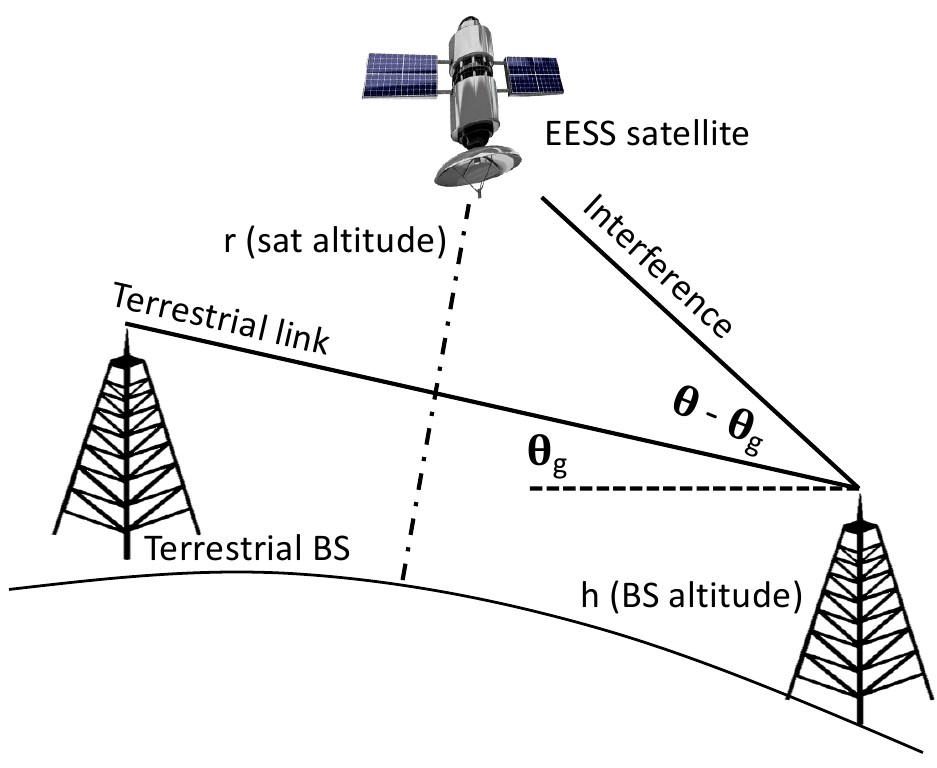}
    \caption{System model}
    \label{fig:system_model}
\end{figure}

We consider the system model shown in Figure~\ref{fig:system_model}: a terrestrial link between two ground stations (e.g., a terrestrial backhaul link~\cite{polese2020toward,petrov2016terahertz}), potentially interfering with a satellite orbiting over the terrestrial deployment area. The ground stations are equipped with directional antennas and can use multiple digital modulations, bandwidths, and carrier frequencies, as shown in Table~\ref{table:scenario}.

For the terrestrial link, the received power \new{$P_{\rm rx,g}(f, d)$} (in dB) can be expressed as a function of the distance $d$ and the carrier frequency $f$:
\begin{equation}
    P_{\rm rx,g}(f, d) = P_{\rm tx,g} - L(f,d) + G_{\rm g, tx, max} + G_{\rm g, rx, max},
\end{equation}
where $P_{tx,g}$ is the transmit power of the ground transmitter, $L$ is the path loss for a terrestrial link, and $G$ is the antenna gain. We assume, without loss of generality, that the transmitter and receiver are aligned and experience the maximum antenna gain.

For the Earth-space link, the received power at the satellite, as a result of the interference from the ground station, can be expressed as (in dB)
\begin{equation}
    P_{\rm rx,s}(f,d,\theta,\theta_{\rm g}) = P_{\rm tx,g} - L(f,d) + G_{\rm tx}(\theta - \theta_{\rm g}) + G_{\rm sat}(\theta - \theta_{\rm g}),
\end{equation}
where $L$ is the path loss for an Earth-space link, $G$ is the antenna gain, $\theta_{\rm g}$ is the tilt of the ground transmitter antenna, and $\theta$ is the elevation angle between the ground transmitter and the satellite\new{, as shown in Figure~\ref{fig:system_model}.}

We model the terrestrial and Earth-space path loss using \gls{itu} path loss models~\cite{itufreespace,ituabsorption,ituatm,iturefractive}, which include the conventional free-space path losses or spreading losses and the molecular absorption losses.
\new{These models are generally considered for the development of \gls{itu} recommendations, and show consistency with other methods (e.g., based on the HITRAN database~\cite{jornet2011thz}) in the frequency range of interest (even though they diverge at higher frequencies)~\cite{ohara2018comment}.}
The free-space attenuation $L_{\mathrm{spr}}$ is due to the expansion of the wave as it propagates through the space and is calculated (in dB) as~\cite{itufreespace}
\begin{equation}
    L_{\rm spr} (f,d) = 92.45+\log{(fd)},
\end{equation}
where $f$ is the frequency of the signal in GHz, and $d$ is the propagation distance in km. For Earth-space slant paths, this distance is
\begin{equation}
    d = \sqrt{R^2\sin^2{\theta}+2Rr+r^2}-R\sin{\theta},
\end{equation}
where $r$ is the satellite altitude, $\theta$ is the elevation angle and $R$ is the Earth radius~\cite{itudistance}.

The molecular absorption losses are due to the interaction of the RF signal with air molecules. As mentioned in Section~\ref{subsec:sensing}, certain frequencies of interest result in the greater resonance with specific molecules, leading to a frequency dependent absorption loss. At frequencies above 100~GHz, the main contributor to absorption loss is water vapor~\cite{jornet2011thz}. Thus, the specific gaseous attenuation \new{at height $h$ is given by $\gamma(h) = \gamma_\text{o}(h) + \gamma_\text{w}(h)$ (dB/km)}, where $\gamma_\text{o}(h)$ and $\gamma_\text{w}(h)$ are the specific attenuation due to dry air and water vapor, respectively~\cite{ituabsorption}. Due to the inherent frequency selective and medium dependent nature of absorption losses, it is dependent both on the frequency of the RF signal, and the characteristics of the medium through which the signal travels (such as temperature, pressure and humidity, and density). For different Earth-space slant paths, the signal traverses different atmospheric heights, which lead to different temperatures and pressures~\cite{ituatm}. Taking into account these factors, the total absorption loss \new{$L_\text{abs}$} is then calculated as
\begin{equation}
    L_\text{abs}=\int_0^r\frac{\gamma(h)}{\sqrt{1-\cos^2\theta(h)}}dh,
\end{equation}
where \new{$\gamma(h)$ is the specific gaseous attenuation at height $h$} and $\theta(h)$ is the local apparent elevation angle at height $h$, which depends on the refractive index at that height \cite{iturefractive}.
Combining these, the total path loss is then
\begin{equation}
    L = L_\text{spr} + L_\text{abs}.
\end{equation}

\begin{table}[t]
    \centering
    \begin{tabular}{ll}
    \toprule
     Parameter & Value  \\\midrule
     Channel model & Based on ITU-R~\cite{itufreespace,ituabsorption,ituatm,iturefractive} \\
     Antenna gain model & Based on ITU-R~\cite{ituantenna} \\
     $G_{\text{g}, i, \text{max}}, i \in \{\text{tx}, \text{rx}\}$ & 40 dBi \\
     $G_{\rm sat}$ &  $\{20, 40\}$ dBi \\
     $P_{\rm tx, g}$ & $\{18.6, 33.4\}$ dBW (see Section~\ref{sec:intf}) \\
     $f_c$ & \{150, 183, 230\} GHz \\
     Noise PSD  & $N_0 = -160$ dBW \\
     Modulation & 16-QAM, 64-QAM, 256-QAM, 1024-QAM\\
     Bandwidth & Based on Figure~\ref{fig:datarates} \\
    \bottomrule
    \end{tabular}
    \caption{System parameters.}
    \label{table:scenario}
\end{table}


\begin{figure*}[t]
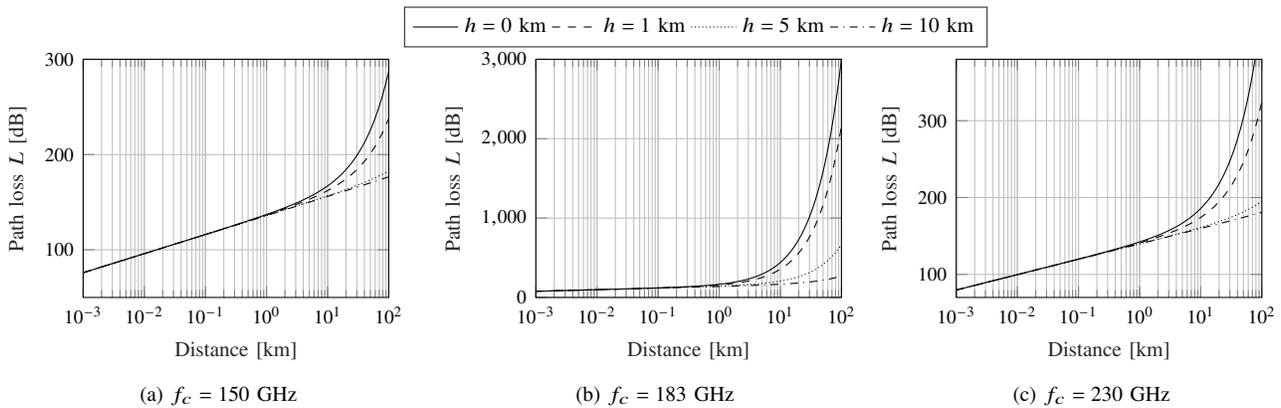

\centering
\ifexttikz
    \tikzsetnextfilename{terrestrial-150-ghz}
\fi
\begin{subfigure}[t]{0.32\textwidth}
    \setlength\fwidth{.7\columnwidth}
  \setlength\fheight{.6\columnwidth}
  \input{Figures/terrestrial-150-ghz.tex}
  \caption{$f_c=150$ GHz}
\end{subfigure}%
\ifexttikz
    \tikzsetnextfilename{terrestrial-183-ghz}
\fi
\begin{subfigure}[t]{0.32\textwidth}
    \setlength\fwidth{.7\columnwidth}
  \setlength\fheight{.6\columnwidth}
  \input{Figures/terrestrial-183-ghz.tex}
  \caption{$f_c=183$ GHz}
  \label{fig:pathloss_ground_183}
\end{subfigure}%
\ifexttikz
    \tikzsetnextfilename{terrestrial-230-ghz}
\fi
\begin{subfigure}[t]{0.32\textwidth}
    \setlength\fwidth{.7\columnwidth}
  \setlength\fheight{.6\columnwidth}
  \input{Figures/terrestrial-230-ghz.tex}
  \caption{$f_c=230$ GHz}
\end{subfigure}%
	\caption{Path loss $L$ as a function of traveled distance for terrestrial links and different altitudes $h$.}
	\label{fig:pathloss_ground}
\end{figure*}

\ifexttikz
    \tikzsetnextfilename{terrestrial-paths-100-275-ghz}
\fi
\begin{figure*}[t]
\centering
  \setlength\fwidth{.8\textwidth}
  \setlength\fheight{.2\textwidth}
  \input{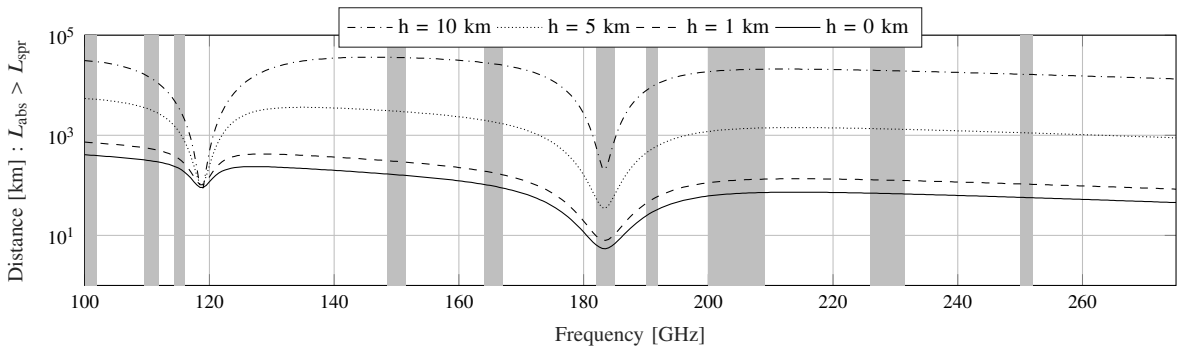}
	\caption{Distance at which $L_{\rm abs}>L_{\rm spr}$. The dark region correspond to RR5.340 bands.}
	\label{fig:pathloss_ground2}
\end{figure*}

The antenna gain model also follows \gls{itu} recommendations. Notably, we consider the model in~\cite{ituantenna} for above 70 GHz, which models the gain as a function of the off-axis angle $\theta$, the carrier frequency $f_c$, and the maximum gain $G_{\rm max}$ (in dB). The \gls{itu} model first computes the antenna diameter \new{$D$} as a function of $G_{\rm max}$ as
\begin{equation}
    D \approx \lambda \cdot 10^{\frac{G_{\rm max} - 7.7}{20}},
\end{equation}
where $\lambda = c / f_c$ is the wavelength. Then, it defines the first-lobe gain $G_1$ as 
\begin{equation}
    G_1 = 2 + 15 \log_{10} \left(\frac{D}{\lambda}\right).
\end{equation}
In the case of $D / \lambda > 100$, the gain \new{$G$ as a function of the angle $\theta$} is computed as
\begin{equation}\label{eq:gain}
    G (\theta) = \begin{cases}
        G_{\rm max} - 2.5\cdot 10^{-3} \left(\frac{D}{\lambda}\theta\right)^2, & 0 \le \theta < \theta_m \\
        G_1, & \theta_m \le \theta < \theta_r \\
        32 - 25\log_{10}(\theta), & \theta_r  \le \theta < 120^{\circ} \\
        - 20, & 120^{\circ} \le \theta < 180^{\circ},
    \end{cases}
\end{equation}
where
\begin{equation}
    \theta_r = 15.85 \left(\frac{D}{\lambda}\right)^{-0.6},
\end{equation}
and
\begin{equation}
    \theta_m = \frac{20 D}{\lambda} \sqrt{G_{\rm max} - G_1}.
\end{equation}
If $D / \lambda \le 100$, instead, the gain is
\begin{equation}\label{eq:gain}
    G (\theta) = \begin{cases}
        G_{\rm max} - 2.5\cdot 10^{-3} \left(\frac{D}{\lambda}\theta\right)^2, & 0 \le \theta < \theta_m \\
        G_1, & \theta_m \le \theta < 100 \frac{\lambda}{D} \\
        52 - 10 \log_{10} \left(\frac{D}{\lambda}\theta\right) - 25 \log_{10} (\theta), & 100 \frac{\lambda}{D} \le \theta < 120^{\circ} \\
        - 10 \log_{10} \left(\frac{D}{\lambda}\theta\right), & 120^{\circ} \le \theta < 180^{\circ}.
    \end{cases}
\end{equation}

\begin{figure*}[t]
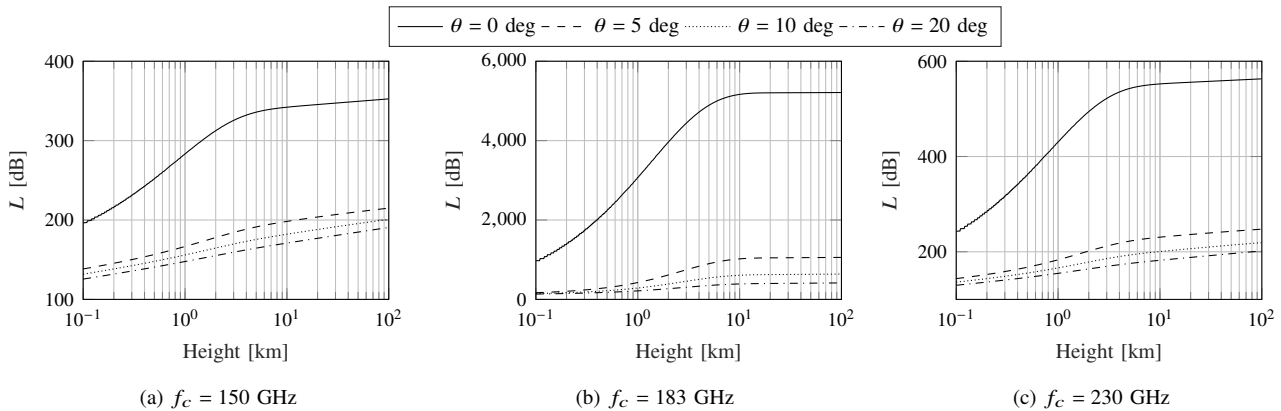

\centering
\ifexttikz
    \tikzsetnextfilename{earth-space-pl-150-ghz}
\fi
\begin{subfigure}[t]{0.32\textwidth}
    \setlength\fwidth{.7\columnwidth}
  \setlength\fheight{.6\columnwidth}
  \input{Figures/earth-space-pl-150-ghz.tex}
  \caption{$f_c=150$ GHz}
\end{subfigure}%
\ifexttikz
    \tikzsetnextfilename{earth-space-pl-183-ghz}
\fi
\begin{subfigure}[t]{0.32\textwidth}
    \setlength\fwidth{.7\columnwidth}
  \setlength\fheight{.6\columnwidth}
  \input{Figures/earth-space-pl-183-ghz.tex}
  \caption{$f_c=183$ GHz}
  \label{fig:pathloss_slant_183}
\end{subfigure}%
\ifexttikz
    \tikzsetnextfilename{earth-space-pl-230-ghz}
\fi
\begin{subfigure}[t]{0.32\textwidth}
    \setlength\fwidth{.7\columnwidth}
  \setlength\fheight{.6\columnwidth}
  \input{Figures/earth-space-pl-230-ghz.tex}
  \caption{$f_c=230$ GHz}
\end{subfigure}%
  \caption{Earth-space path loss as a function of the satellite height, for different elevation angles $\theta$.}
  \label{fig:pathloss_slant}
\end{figure*}

The \gls{snr} $\Gamma = P_{\text{rx},i} / (N_0B),\;i \in \{\rm g, s\}$ is computed using a noise \gls{psd} $N_0 = -160$ dBW. 
\new{Here, we consider a noise model based on empirical measurements on the receivers in the TeraNova platform~\cite{SEN2020107370}, where the electronic noise temperature of the system dominates the noise temperature due to molecular absorption and the overall amplitude noise follows a Gaussian distribution.}
For $B$, we consider the bandwidth that is required to achieve a 1 Tbps data rate with a certain modulation and coding scheme, as shown in Figs.~\ref{fig:datarates_mmimo} and~\ref{fig:datarates_mimo} (Section~\ref{sec:stakeholders}). We then utilize the \gls{snr} to estimate the error rate probability for specific digital modulation and coding rates~\cite{john2008digital}.

\subsection{Path Loss Analysis}
\label{sec:pathloss}

We first analyze the path loss for the terrestrial and Earth-space links.

\subsubsection{Terrestrial links}
Figure~\ref{fig:pathloss_ground} shows the path loss at three different carrier frequencies, for a link between two ground stations at different altitudes. The path loss curves have a typical L-shaped trend, as the spreading loss increases logarithmically with the distance (thus linearly with respect to the logarithmic plots in Figure~\ref{fig:pathloss_ground}), while the absorption loss is linear with the distance (thus exponential in the plots of Figure~\ref{fig:pathloss_ground}). This can be seen also by comparing the path loss at different altitudes. At higher altitudes, the atmosphere is less dense; the absorption loss reduces accordingly. As shown in Figure~\ref{fig:pathloss_ground}, the path loss for $h=10$ km maintains a logarithmic trend with the distance, while the L-shape is more evident for the path loss at $h=0$ km.
Finally, the path loss is similar \new{for $f_c = 150$~GHz and for $f_c=230$}~GHz, with the latter having a slightly higher spreading loss. 
Instead, when considering $f_c=183$ GHz (Figure~\ref{fig:pathloss_ground_183}) the impact of the absorption is dominant, as this frequency corresponds with a water vapour absorption peak. The path loss due to absorption is again reduced at higher altitudes, due to a less dense atmosphere.

\begin{figure}[t]
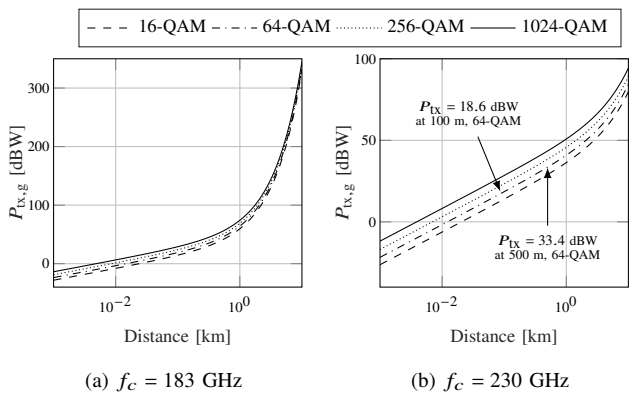

\centering
\ifexttikz
    \tikzsetnextfilename{tbps-ptx-183}
\fi
\begin{subfigure}[t]{0.49\columnwidth}
    \setlength\fwidth{.8\columnwidth}
  \setlength\fheight{.7\columnwidth}
  \input{Figures/tbps-ptx-183.tex}
  \caption{$f_c = 183$ GHz}  
  \label{fig:tbps-ptx-183}
\end{subfigure}%
\ifexttikz
    \tikzsetnextfilename{tbps-ptx-230}
\fi
\begin{subfigure}[t]{0.49\columnwidth}
    \setlength\fwidth{.8\columnwidth}
  \setlength\fheight{.7\columnwidth}
  \input{Figures/tbps-ptx-230.tex}
  \caption{$f_c = 230$ GHz}  
  \label{fig:tbps-ptx-230}
\end{subfigure}
  \caption{Transmit power $P_{\rm tx,g}$ to achieve a 1 Tbps data rate with a bit error rate smaller than $10^{-5}$, for different modulations and as a function of the distance between the terrestrial transmitter and receiver.}
  \label{fig:tbps-ptx}
\end{figure}

To further investigate the interplay between spreading and absorption loss, Figure~\ref{fig:pathloss_ground2} reports the distances (in km) at which the absorption loss $L_{\rm abs}$ becomes larger than the spreading loss $L_{\rm spr}$, for different frequencies and altitudes of the link. These results indicate that---outside absorption peaks---the contribution of $L_{\rm abs}$ is smaller than that of $L_{\rm spr}$ even for tens of km, even for the dense atmosphere at $h=0$ km, demystifying the claim that communications in the above 100~GHz band are limited by the absorption phenomena. 

\subsubsection{Earth-space Slant Paths} 
\label{sec:earth-space-slant}

The different composition and density of the Earth's atmosphere at different altitudes has an impact on the path loss trend for Earth-space paths. 
Figure~\ref{fig:pathloss_slant} reports the path loss $L$ for three different frequencies, as a function of the satellite altitude and the elevation angle between the ground station and the satellite.

The curves exhibit a reverse L-shaped trend, especially for an elevation of 0 degrees. This shows that there exists two regimes for the path loss. Below a certain altitude threshold, which is at most 10 km for the elevation of 9 degrees and $f_c = 183$ GHz, the path loss is affected by the combined spreading and absorption losses. This leads to a much higher path loss for $f_c = 183$ GHz (Figure~\ref{fig:pathloss_slant_183}) than $f_c = 150$ GHz and $230$ GHz. Above this threshold, however, the less dense atmosphere leads to a reduced impact of the absorption loss $L_{\rm abs}$, with the dominant contribution given by the spreading loss $L_{\rm spr}$. This behavior is more evident for configurations with the satellite at a lower elevation, as the signal has to travel through a thicker portion of the Earth atmosphere. The higher path loss for lower elevation angles is key to reducing interference from ground stations to satellites, as we discuss in Section~\ref{sec:intf}.


\subsection{Link Budget and \gls{rfi} Analysis}
\label{sec:intf}

To identify a meaningful transmit power value for the ground station in the interference analysis, we analyze the link budget for a terrestrial link that supports a data rate of 1 Tbps with a \gls{ber} smaller than $10^{-5}$. We consider the system model described in Section~\ref{sec:system_model}, with the parameters from Table~\ref{table:scenario}, and the bandwidth for different modulations shown in Figure~\ref{fig:datarates_mimo}. 

Figure~\ref{fig:tbps-ptx} reports the transmit power $P_{\rm tx,g}$ that satisfies the required constraints for $f_c = 183$ GHz and $f_c=230$ GHz, and four different QAM modulation orders.
\new{Considering that the number of \gls{mimo} streams supported by the link depends on the diversity of the physical channel and not on the wireless link design itself, we compute the target $P_{\rm tx,g}$ by considering the average of the transmit power required with 2, 4, or 8 \gls{mimo} channels}.


As expected, the impact of the higher absorption loss at 183~GHz translates into a higher required $P_{\rm tx,g}$ than at 230~GHz, in particular for transmitter-receiver distances larger than 1 km. Additionally, high-order modulations need a higher $P_{\rm tx,g}$ than low-order modulations, with an average 14.46 dB gap between 16-QAM and 1024-QAM. This disparity exists in spite of the reduced bandwidth requirements with high-order modulations to achieve the target rate of 1~Tbps.

In the following paragraphs, without loss of generality, we consider two arbitrary reference values for $P_{\rm tx,g}$, i.e., $P_{\rm tx,g} = 18.6$ dBW, which enables a date rate of 1 Tbps with 64-QAM at 100~m, and $P_{\rm tx,g} = 33.4$ dBW, which corresponds to 1~Tbps data rate at 500~m, again with 64-QAM. These distances are typical representations of a backhaul link in a dense, urban environment. \new{While we acknowledge that such transmit power levels (i.e., 72 W and 2.18 kW) are not feasible at the moment in above-100-GHz devices, they allow us to speculate on interference levels that orbiting satellites may experience in a typical 6G deployment. Similar values are reported for transmit power and effective isotropic radiated power in current sub-6 GHz and lower \gls{mmwave} backhaul deployments~\cite{ge20145g,adare2021uplink}.}

\begin{table}[t]
    \centering
    \begin{tabularx}{\linewidth}{>{\raggedright\arraybackslash\hsize=.95\hsize}X>{\raggedright\arraybackslash\hsize=1\hsize}X>{\raggedright\arraybackslash\hsize=.075\hsize}X>{\raggedright\arraybackslash\hsize=1.8\hsize}X} \toprule
         Band [GHz] & Maximum Interference Level [dBW] & \multicolumn{2}{l}{Lowest Satellite Altitude [km]} \\\midrule
         115.25-122.25 & $-166$ (N), $-189$ (L) & 705 & Aura (NASA)~\cite{schoeberl2006aura}  \\
         148.5-151.5 & $-159$ (N), $-189$ (L) & 705 & Aqua (NASA)~\cite{parkinson2003aqua} \\
         155.5-158.5 & $-163$ (N,C) & 817 & MetOp (EUMETSAT)~\cite{edwards2000metop} \\
         164-167 & $-163$ (N,C), $-189$~(L) & 407 & GPM Core Observatory (NASA/JAXA)~\cite{skofronick2017global} \\
         174.8–191.8 & $-163$ (N,C), $-189$~(L) & 407 & GPM Core Observatory (NASA/JAXA)~\cite{skofronick2017global} \\
         226-231.5 & $-160$ (N,C), $-194$~(L) & 705 & Aura (NASA)~\cite{schoeberl2006aura}  \\
         235-238 & $-194$ (L) & 705 & Aura (NASA)~\cite{schoeberl2006aura}  \\
         \bottomrule
    \end{tabularx}
    \caption{Passive sensing satellites in different bands above 100~GHz, with the maximum interference level set in~\cite{itu2012perf}. N stands for Nadir scan, C for conic, L for limb.}
    \label{table:satellites}
\end{table}

Table~\ref{table:satellites} lists the satellite which, for a given frequency band, occupies the lowest altitude orbit. The maximum permissible interference level is defined in~\cite{itu2012perf}. The \gls{rfi} level changes according to whether the scan is performed in Nadir, Conic, or Limb mode, i.e., according to the orientation and angle between the ground \gls{rfi} emitter and the antenna of the satellite under consideration. \new{With a Nadir scan, the satellite sensor and/or antenna points to the ground directly below the satellite (i.e., the Nadir direction). With a Limb scan, the sensor and/or antenna points to the edge of the Earth as seen by the satellite. With a Conic scan, the satellite can sense in different directions (cones, or beams) over time.}
The \gls{rfi} threshold is also defined independently on the specific scientific mission and the altitude at which satellites orbit. Therefore, we consider this as a reference for harmful \gls{rfi} in the next paragraphs.

\begin{figure}[t]
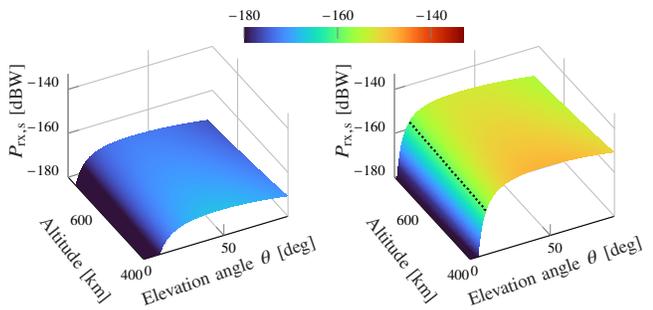
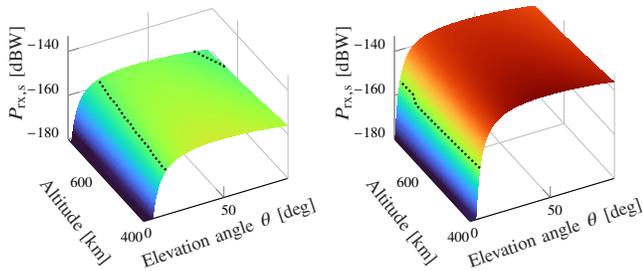

\centering
\ifexttikz
    \tikzsetnextfilename{interference-power-186-20}
\fi
\begin{subfigure}[t]{0.49\columnwidth}
    \setlength\fwidth{.8\columnwidth}
  \setlength\fheight{.7\columnwidth}
  \input{Figures/interference-power-186-20.tex}
  \caption{$P_{\rm tx,g} = 18.6$ dBW, $G_{\rm sat} = 20$ dBi}  
  \label{fig:interference-power-186-20}
\end{subfigure}%
\ifexttikz
    \tikzsetnextfilename{interference-power-186-40}
\fi%
\begin{subfigure}[t]{0.49\columnwidth}
    \setlength\fwidth{.8\columnwidth}
  \setlength\fheight{.7\columnwidth}
  \input{Figures/interference-power-186-40.tex}
  \caption{$P_{\rm tx,g} = 18.6$ dBW, $G_{\rm sat} = 40$ dBi}  
  \label{fig:interference-power-186-40}
\end{subfigure}\\
\ifexttikz
    \tikzsetnextfilename{interference-power-334-20}
\fi%
\begin{subfigure}[t]{0.49\columnwidth}
    \setlength\fwidth{.8\columnwidth}
  \setlength\fheight{.7\columnwidth}
  \input{Figures/interference-power-334-20.tex}
  \caption{$P_{\rm tx,g} = 33.4$ dBW, $G_{\rm sat} = 20$ dBi}  
  \label{fig:interference-power-334-20}
\end{subfigure}%
\ifexttikz
    \tikzsetnextfilename{interference-power-334-40}
\fi%
\begin{subfigure}[t]{0.49\columnwidth}
    \setlength\fwidth{.8\columnwidth}
  \setlength\fheight{.7\columnwidth}
  \input{Figures/interference-power-334-40.tex}
  \caption{$P_{\rm tx,g} = 33.4$ dBW, $G_{\rm sat} = 40$ dBi}  
  \label{fig:interference-power-334-40}
\end{subfigure}
\caption{\gls{rfi} power $P_{\rm rx,s}$ at a satellite, as a function of satellite altitude and elevation angle with respect to the ground station, for different values of $P_{\rm tx,g}$ and $G_{\rm sat}$. The dotted line represents the combinations of altitude and elevation angle values that lead to $P_{\rm rx,s} > -160$ dBW (see Table~\ref{table:satellites}). $f_c = 230$ GHz, $\theta_{\rm g} = 0$ degrees.}
\label{fig:interference}
\end{figure}

Figure~\ref{fig:interference} analyzes the \gls{rfi} for satellites performing a scan at $f_c = 230$ GHz. \new{We assume Nadir/Conic operations, for which the corresponding \gls{rfi} threshold is $-160$ dBW.} Herein, the ground station antenna has a tilt $\theta_{\rm g} = 0$ degrees.  
Additionally, the satellite altitude $r$ ranges from 407 to 720~km, and we consider the two power levels identified in Figure~\ref{fig:tbps-ptx-230}, two different (and constant) gain $G_{\rm sat}$ values for the satellite antenna (i.e., 20 dB and 40 dB), and an elevation angle $\theta$ between the satellite and the ground station that varies between 0 and 90 degrees. 

With the lowest transmit power and satellite antenna gain (i.e., 18.6~dBW and 20~dBi, respectively), the interference never exceeds the harmful \gls{rfi} threshold, as shown in Figure~\ref{fig:interference-power-186-20}. However, when the gain of the satellite antenna increases to 40~dBi, as shown in Figs.~\ref{fig:interference-power-186-40} and~\ref{fig:interference-power-334-40}, the interference threshold is exceeded when the satellite orbits above a certain threshold elevation angle, $\theta_{th}$ (represented by the dotted line). The value of $\theta_{th}$ increases with the satellite altitude. For $P_{\rm tx,g} = 18.6$ dBW (Figure~\ref{fig:interference-power-186-40}), it ranges from 9.73 degrees at 400 km to 12.26 degrees at 720 km. For $P_{\rm tx,g} = 33.4$ dBW (Figure~\ref{fig:interference-power-334-40}), it varies from 6 degrees for 400 km to 6.64 degrees for 720 km. 

Finally, for $P_{\rm tx,g} = 33.4$ dBW and $G_{\rm sat} = 20$ dBi (Figure~\ref{fig:interference-power-334-20}), satellites at lower altitudes (i.e., below 645~km) experience harmful \gls{rfi} above the ITU-defined threshold for angles above 13.89 degrees (for 400 km) and 19.3 degrees (for 645~km). The trend is different for higher-altitude satellites, for which there also exists a region at high elevation angles in which the interference is below $-160$ dBW. Namely, satellites orbiting above 645~km experience \gls{rfi} above threshold in for elevation angles $\theta$ between 19.3 and 89 degrees (for $r=645$ km) and between 21.74 and 81.54 degrees (for $r=720$ km). 

\ifexttikz
    \tikzsetnextfilename{interference-gain-vs-loss}
\fi
\begin{figure}[b]
\centering
  \setlength\fwidth{.7\columnwidth}
  \setlength\fheight{.35\columnwidth}
  \input{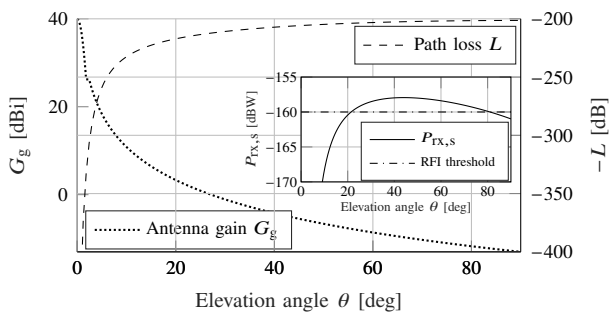}
  \caption{Antenna gain $G_{\rm g}$, path gain $-L$, and received power $P_{\rm rx, s}$ for different elevation angles, satellite height $r=700$ km, $f_c = 230$ GHz, $G_{\rm sat}=20$ dBi, $P_{\rm tx, g} = 33.4$ dBW.}
  \label{fig:gain-angle}
\end{figure}

To better understand the elevation-dependent \gls{rfi} trend, we present, in Figure~\ref{fig:gain-angle}, the ground station antenna gain $G_{\rm g}$ (dotted, left axis), the path gain $-L$ (dashed, right axis), and the overall received \gls{rfi} $P_{\rm rx, s}$ for $f_c = 230$ GHz, $P_{\rm tx, g} = 33.4$ dBW, and $G_{\rm sat} = 20$ dBi, and $r=700$ km. It is worth noting that for the same satellite altitude $r$, the distance between the satellite and a particular ground station is greater for lower elevation angles, i.e., when the satellite is closer to the horizon. Additionally, the signal has to travel through a thicker portion of the Earth's atmosphere, further increasing the contribution of the absorption loss. Therefore, even if the antenna gain of the ground station antenna is maximum at $\theta = \theta_{\rm g} = 0$ degrees, the high path loss leads to a \gls{rfi} below the defined threshold. The path loss, however, decreases faster than the gain as a function of the elevation angle $\theta$, as shown in Figure~\ref{fig:gain-angle}. This leads to an increase in $P_{\rm rx,s}$, which---in this specific case---exceeds the \gls{itu} threshold for $\theta > 21$ degrees. Finally, the path loss flattens for elevation angles above 50 degrees, while the antenna gain decreases logarithmically with $\theta$. Therefore, there can be combinations of path loss, transmit power, and antenna gain values that lead to \gls{rfi} being below the \gls{itu} threshold for high elevation angles (i.e., for the satellite orbiting above the ground station).

\begin{figure}[t]
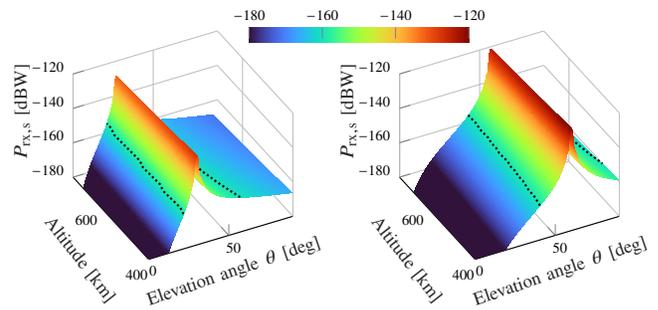

\centering
\ifexttikz
    \tikzsetnextfilename{interference-power-186-20-tilt-30}
\fi
\begin{subfigure}[t]{0.49\columnwidth}
    \setlength\fwidth{.8\columnwidth}
  \setlength\fheight{.7\columnwidth}
  \input{Figures/interference-power-186-20-tilt-30.tex}
  \caption{$\theta_{\rm g}=30$ degrees.}  
  \label{fig:interference-tilt-30}
\end{subfigure}%
\ifexttikz
    \tikzsetnextfilename{interference-power-186-20-tilt-60}
\fi%
\begin{subfigure}[t]{0.49\columnwidth}
    \setlength\fwidth{.8\columnwidth}
  \setlength\fheight{.7\columnwidth}
  \input{Figures/interference-power-186-20-tilt-60.tex}
  \caption{$\theta_{\rm g}=60$ degrees.}  
  \label{fig:interference-tilt-60}
\end{subfigure}
\caption{\gls{rfi} power $P_{\rm rx,s}$ at a satellite, as a function of satellite altitude and elevation angle with respect to the ground station, for different values of the ground antenna tilt. The dotted line represents the combinations of altitude and elevation angle values that lead to $P_{\rm rx,s} > -160$ dBW (see Table~\ref{table:satellites}). $f_c = 230$ GHz, $P_{\rm tx,g} = 18.6$ dBW, $G_{\rm sat} = 20$ dBW.}
\label{fig:interference-tilt}
\end{figure}

We also consider the case in which the ground station antenna points toward the sky with a positive tilt $\theta_{\rm g}$. This is a typical representative of 6G non-terrestrial use cases, such as satellite-assisted cellular networks~\cite{rinaldi2020non}, the space Internet~\cite{alqaraghuli2021performance}, and drones and \gls{uav} aerial networks~\cite{wan2020space,bertizzolo2020live}. Figure~\ref{fig:interference-tilt} reports the received power at the satellite for $P_{\rm tx, g} = 18.6$ dBW and $G_{\rm sat} = 20$ dBi, and the antenna on the ground tilted by $\theta_{\rm g} = 30$ degrees and $\theta_{\rm g} = 60$ degrees. Notice that with $\theta_{\rm g} = 0$, the \gls{rfi} for this configuration of transmit power and satellite gain does not exceed $-160$ dBW (Figure~\ref{fig:interference-power-186-20}). \new{Our analysis shows that there exists a tilt angle $\theta_{\rm g, th} = 7.2$ for which combinations of $\theta_{\rm g} \ge \theta_{\rm g, th}$, $\theta$, and $r$ lead to \gls{rfi} above threshold.}
%
In particular, the satellite receives \gls{rfi} as high as $-127$ dBW for $\theta_{\rm g} = 30$ degrees (Figure~\ref{fig:interference-tilt-30}), and $-119$ dBW for $\theta_{\rm g} = 60$ degrees (Figure~\ref{fig:interference-tilt-60}).

These analyses indicate that directional antennas can come with mixed blessings: while the combination of directivity and high atmospheric absorption helps reducing \gls{rfi} in some scenarios (low satellite elevation), the same can lead to extremely high \gls{rfi} for higher satellite elevation and non-terrestrial communications scenarios.
Finally, while we consider ground-to-satellite \gls{rfi} herein, similar insights can be derived when investigating interference from aerial sources toward ground passive scientific infrastructure, e.g., for radio astronomy. 
Indeed, while terrestrial radio astronomy sites are generally isolated and in remote locations, sidelobes (or, more rarely, main lobes) from aerial sources have the potential to generate harmful interference to passive terrestrial stations\new{, as discussed in Section~\ref{sec:current_regulations}}.\footnote{Notice that large satellite constellations also have the potential to affect ground based \emph{optical} telescopes used for astronomy and planetary defense\new{, i.e., to track satellites and near-Earth asteroids. However, }it has been found that the impact from Low Earth Orbit satellites is quite manageable, as their reflected sunlight is rarely bright enough to cause major problems, except when they are very close to the horizon where these telescopes rarely operate~\cite{eso2020report}.}

\section{Spectrum Sharing Techniques Above 100~GHz --- Challenges and Opportunities}
\label{sec:sharing}

The results of Section~\ref{sec:interference} highlight that there exist scenarios in which active transmissions can create interference toward passive users, despite the directional nature of the signals and the high absorption and spread loss. \newrev{At the same time, however, they suggest that in-band sharing in the spectrum above 100 GHz may be possible, thanks to the combination of the molecular and spreading path losses and the directional transmissions.} Therefore, to enable dynamic spectrum sharing solutions, it becomes necessary to identify some technological enablers that allow \newrev{network operations in a regime where (i) interference is reduced or eliminated, or (ii) shared usage is enabled on a spatial, time, or frequency/band basis.} The next paragraphs discuss the technological challenges to spectrum sharing above 100~GHz, with prospective solutions that span from devices and RF circuits engineering (antenna solutions, innovative intelligent surfaces) to the design of communication and sensing stacks. \newrev{Note that some of these technologies are is use---or can be used---also in other frequency bands, but they often require adaptations or addressing specific engineering challenges to be effective or feasible above 100 GHz, as we discuss in the \emph{Research challenges} paragraphs of this section.}

\begin{figure*}[h!]
\centering
\begin{subfigure}[t]{0.32\textwidth}
\centering
    \includegraphics[width=0.75\linewidth]{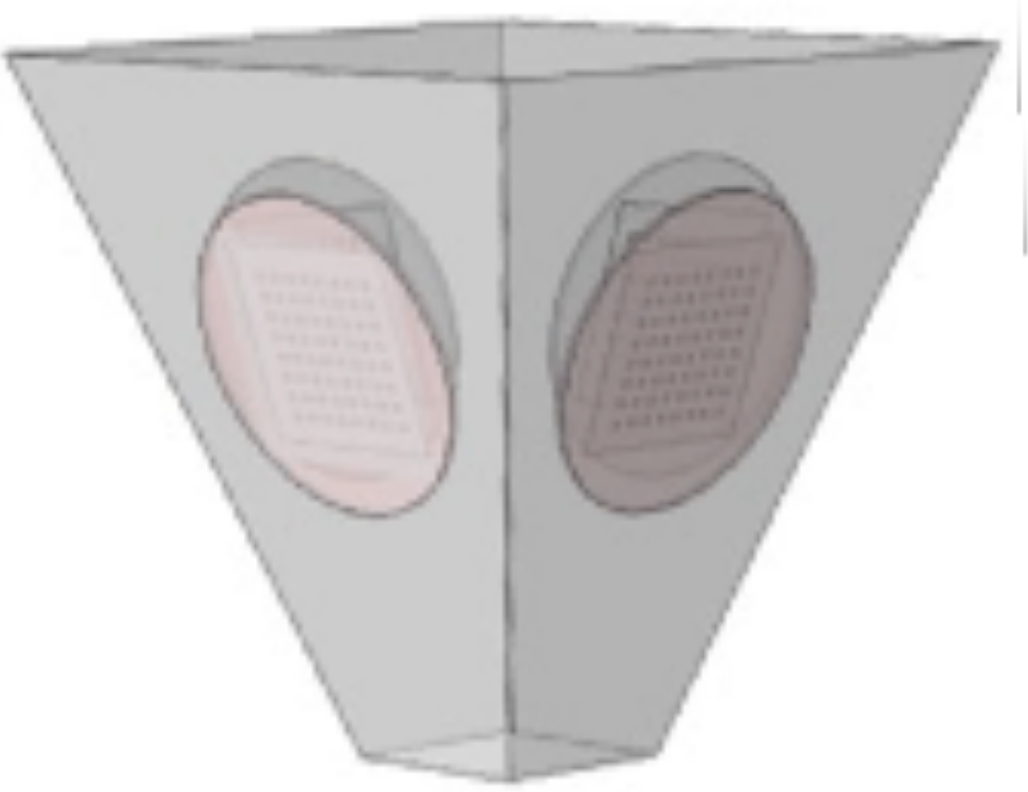}
    \caption{Inverse pyramid design for a terrestrial BS can minimize interference with other users of interest above the BS height.}
    \label{fig:device_design_pyramid}
\end{subfigure}\hfill
\begin{subfigure}[t]{0.32\textwidth}
\centering
    \includegraphics[width=0.75\linewidth]{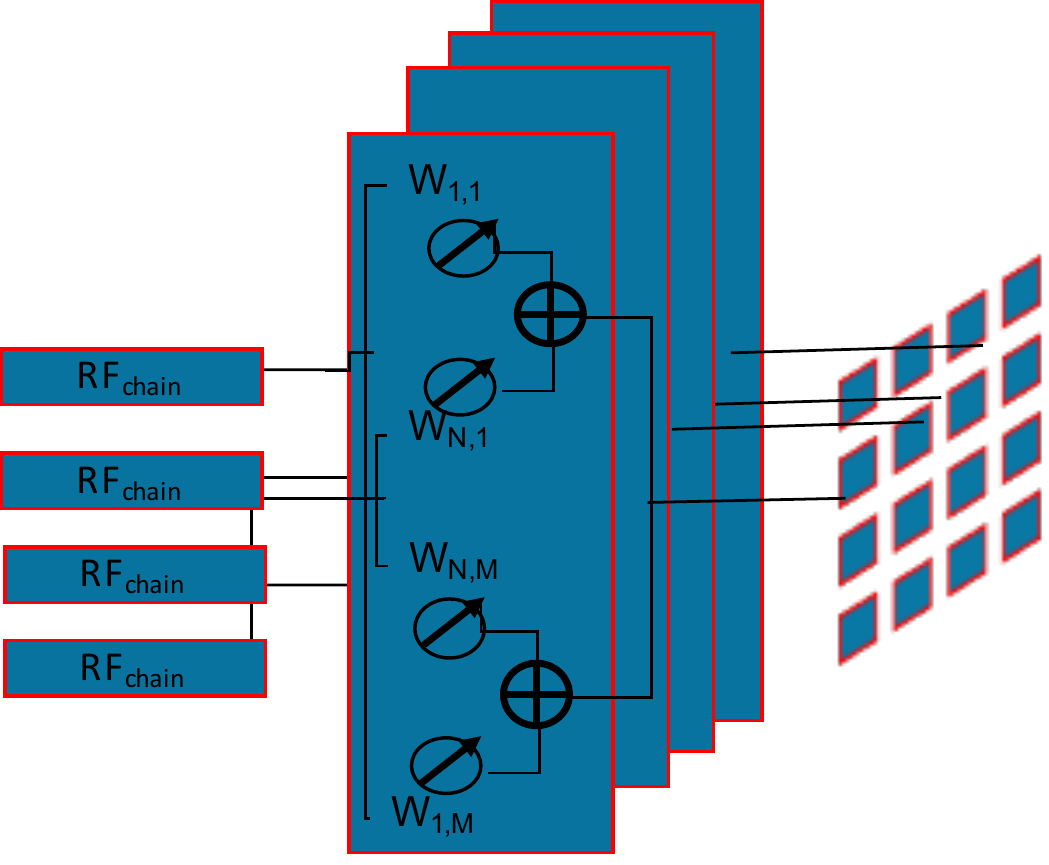}
    \caption{Hybrid or fully digital array architectures can be utilized to implement non-uniform amplitude arrays for reduced sidelobe strength~\cite{balanis2016antenna}.}
    \label{fig:device_design_array}
\end{subfigure}\hfill
\begin{subfigure}[t]{0.32\textwidth}
\centering
    \includegraphics[width=0.4\linewidth]{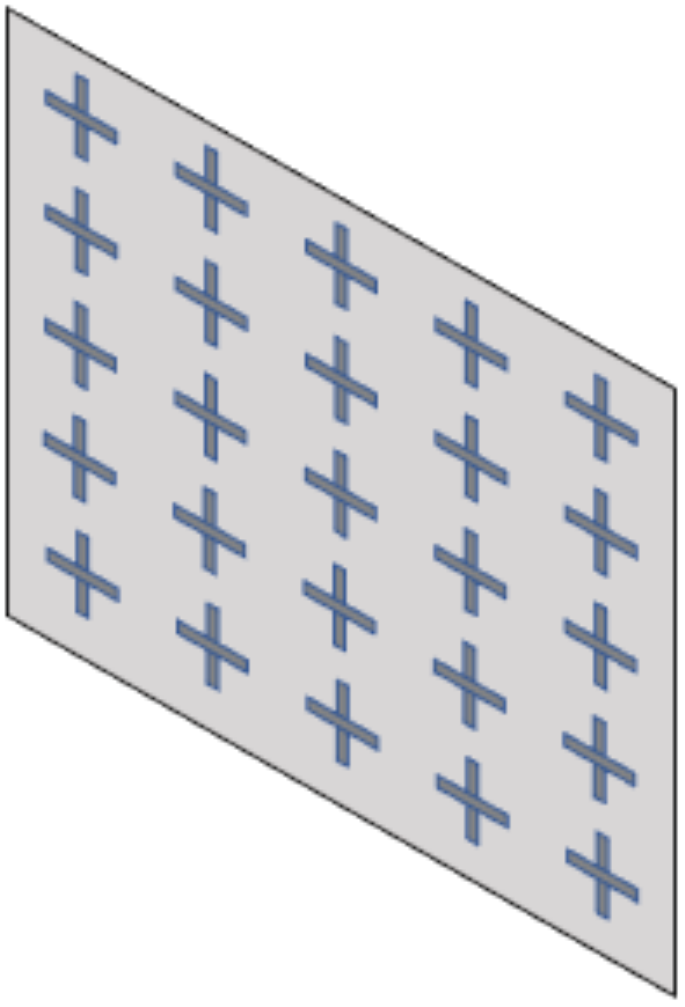}
    \caption{An FSS can be incorporated to act as a passive aid to block frequencies of interest from extremely sensitive directions, for example through a dome structure as discussed in~\cite{jamnejad2002array}.}
    \label{fig:device_design_fss}
\end{subfigure}

	\caption {Device configurations to minimize undesired \gls{rfi}. 
 }
	\label{fig:device_design}
\end{figure*}

\subsection{Challenges to Spectrum Sharing Solutions Between Active and Passive Users Above 100~GHz}

Spectrum sharing has been widely studied in the sub-6 GHz domain, as mentioned in Section~\ref{sec:introduction}~\cite{zhang2017survey,voicu2019survey,hu2018full,HAN201653}. A significant effort has been put into identifying spectrum sharing solutions for example in the context of \gls{ran} sharing among multiple 5G operators~\cite{zhang2017survey}. So far, however, most of the research has focused on sharing among active users and, in particular, between communications users. As mentioned above, several of the techniques investigated for the bands below 6~GHz and active users sharing do not necessarily apply to sharing among active and passive users above 100~GHz, or require solving significant engineering challenges.

For example, one of the main areas of research is cognitive radios~\cite{hu2018full}. In this approach, dynamic spectrum access of secondary users over resources primarily destined for other users is enabled with a two-step approach. 
The first step involves sensing or collecting information on spectrum usage, thereby identifying unused portions of the spectrum and/or patterns of usage of the primary users. 
The second step involves secondary-user transmission. \new{This can happen on unused resources (interweave mode) or on the resources already occupied by the primary (overlay mode), possibly with some coordination to satisfy predefined interference constraints}. 
Notably, cognitive radios have been popular in the context of improving the usage of TV white space~\cite{fitch2011wireless}. However, the paradigm of cognitive radio cannot be applied directly to active-passive sharing above 100~GHz. The ``sensing'' step is challenged by several conditions. The spectrum access by passive users cannot be sensed. The increased directivity and large bandwidth requirements of active-user transmissions make detection more complex. Along these lines, the scale (in the number of devices) envisioned for certain above-100-GHz applications~\cite{polese2020toward}\footnote{\new{Projections from literature and market analysis point to an increase from $10^6$ to $10^7$ devices per km$^2$ when transitioning from 5G to 6G~\cite{zhang20196g}.}} further complicates cognitive radio systems, \new{as the overhead for coordination increases and the amount of unused resources for interweave operations decreases}. Furthermore, concurrent cognitive transmissions (e.g., as overlay) may not be tolerated by some passive sensing users if further precautions are not applied to avoid exceeding harmful \gls{rfi} thresholds, as discussed in Section~\ref{sec:interference}.

Other techniques that are broadly considered enablers of spectrum sharing at lower frequencies and for active users cannot be directly applied in the context of active-passive sharing above 100~GHz. \gls{noma}~\cite{dai2018survey}, \new{which enables code-based or power-based sharing} of the same time and frequency resources among multiple users, has been designed for the coexistence of multiple active users. Similarly, sharing across licensed and unlicensed bands does not apply when passive users are at stake~\cite{zhang2019spectrum}.
While~\cite{ramadan2017new} and~\cite{cho2020coexistence} consider coexistence between communication users and passive \gls{ras} and \gls{eess} services, these are for frequency bands below 100~GHz.

\begin{figure}[t]
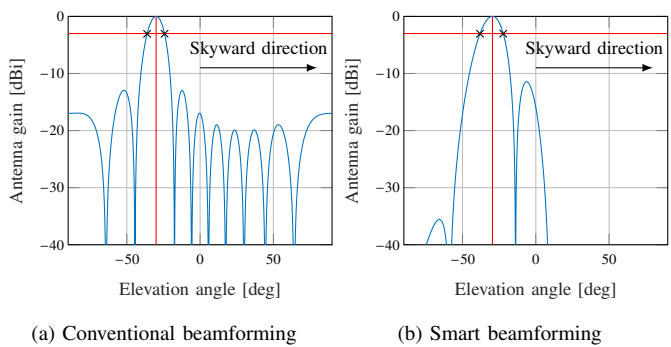

\ifexttikz
    \tikzsetnextfilename{no-control}
\fi
\begin{subfigure}[t]{0.49\columnwidth}
    \centering
    \setlength\fwidth{.85\columnwidth}
    \setlength\fheight{.7\columnwidth}
    \input{Figures/no-control.tex}
    \caption{Conventional beamforming}
    \label{fig:no_control}    
\end{subfigure}
\ifexttikz
    \tikzsetnextfilename{total-control}
\fi
\begin{subfigure}[t]{0.49\columnwidth}
    \centering
    \setlength\fwidth{.85\columnwidth}
    \setlength\fheight{.7\columnwidth}
    \input{Figures/total-control.tex}
    \caption{Smart beamforming}
    \label{fig:total_control}    
\end{subfigure}
    \caption{The utilization of active suppression on the skyward directions during the array synthesis can suppress unwanted emissions. In both cases, the desired angle of steering is \ang{30} below the horizon.}
    \label{fig:my_label}
\end{figure}


An additional challenge is represented by the heterogeneity between the different technologies at these frequencies~\cite{voicu2019survey}, which require either custom solutions or coordination among different standardization or technology development forums. Such coordination, however, has been shown to be previously feasible in the sub-6 GHz context, with the notable example of the \gls{cbrs} band. This portion of the spectrum, which was exclusively allocated to military use (e.g., US Navy radars and satellite links), is now operated according to a shared usage pattern, facilitated by an inter-technology database that keeps track of spectrum usage and grants access with high granularity. 

Overall, it is clear that there exists a need to carefully design and re-think enabling technologies and strategies to allow sharing between active and passive users in the 100~GHz bands. In addition to accounting for the presence of passive users, spectrum sharing above 100~GHz should also factor in and exploit the unique characteristics of the propagation of \gls{rf} signals in these bands~\cite{marcus2023millimeter}. To this end, in the next paragraphs, we discuss and review techniques that can be applied for effective and safe active/passive sharing at frequencies above 100~GHz.


\subsection{Solutions enabled by (sub) Terahertz Devices and Wave Propagation}




To effectively communicate at frequencies above 100~GHz, highly efficient, directional antennas are required. In such a setup, unwanted interference toward sensitive devices outside the main coverage region manifests as leaking power in the form of side-lobes. In directional antennas as well as uniform amplitude antenna arrays, the more directional the mainlobe, the greater the dominant \gls{sll}, or the effective power contained within the dominant sidelobes~\cite{balanis2016antenna}. Such a manifestation of the sidelobes can be seen for a typical high density MIMO array (8 $\times$ 8) in Figure~\ref{fig:no_control}. It is seen that, while the mainlobe can be extremely directional, the major sidelobe strength can be as much as $-13$~dB in comparison. Thus, it is necessary to consider the entire radiation pattern of the array/antenna, and develop working objectives to truncate such \gls{sll} which contribute to interference concerns. 

\textbf{Inverse Pyramidal Arrays.} Above 100~GHz, highly directive radiation as well as significant power output is expected for \glspl{bs} supporting massive MIMO. Thus, the main objective of device design can be tailored towards significant control over the \gls{sll} in these \glspl{bs}. An intuitive method of mitigating interference in unwanted directions is to have a physical configuration within which the unwanted directions are generally excluded from the radiation pattern. The inverse pyramidal array design, an example of which is presented in Figure~\ref{fig:device_design_pyramid}, acts as a direct complement to the pyramidal array design~\cite{khalifa2007geometric}. \new{In such a setup, the geometrical configuration of the array favors scan angles below the horizon, thereby minimizing the power leakage above the BS itself.} A four-sided pyramidal array, similar to that in Figure~\ref{fig:device_design_pyramid}, is known to provide maximum coverage efficiency, thereby reducing deployment resources~\cite{khalifa2007geometric}. Each side of the pyramid comprises an antenna array that serves a particular sector, for full coverage. 

\new{\textit{Research challenges:}} \new{It is important to note that this design can be used only in deployments with \glspl{bs} altitude higher than that of the connected users. In addition, there are multiple research challenges related to building dynamic beamforming antenna arrays above 100~GHz, as we describe together with the next point.}

\textbf{Active \gls{sll} Suppression.} As mentioned in Section~\ref{sec:interference}, 6G \glspl{bs} may be required to provide coverage in aerial scenarios. Here, the \gls{bs} needs to serve users not just at lower or at equivalent altitudes to itself (e.g., other \glspl{bs} and \glspl{ue}), but also to users above in altitude, for example, \glspl{uav}. Thus, in the full 3D coverage space, it becomes necessary to consider more active \gls{sll} suppression techniques. 
In conventional arrays, zero-forcing beamforming is considered a convenient technique, with which \new{different receivers can experience orthogonal channels}. \new{In this case, the goal of the beamforming algorithm is to ensure that no signal is received in a target direction, at the cost of less control on the beam shape everywhere else.}
%
%
However, with fully digital or even hybrid beamforming architectures as found in current massive MIMO 5G devices~\cite{molisch2017hybrid}, more robust beamforming techniques can be implemented. 
We present an example of a hybrid beamforming architecture that could be exploited to control interference in Figure~\ref{fig:device_design_array}, with the corresponding potential far-field radiation response shown in Figure~\ref{fig:total_control}. 
With knowledge of critical interference directions (where sensitive devices may be located) the array response can be tailored through interference suppression algorithms as in~\cite{varade2009robust}, to ensure complete suppression of leakage in critical directions, \new{while still providing coverage to users that require service in other angular regions}. 
As shown in Figure~\ref{fig:total_control}, the mainlobe can be preserved while the more problematic sidelobes can be truncated greatly. 

In addition to novel array architectures, another breakthrough that is extremely important in the above 100~GHz range is the rise of intelligent reflecting surfaces (IRSs)~\cite{di2019smart}. Although the fundamental principles of IRSs are valid across the spectrum, these are increasingly relevant for the (sub-)THz band, as the greatly simplified design helps to reduce the constraints on increasing the size of the array and in its deployment~\cite{monroe2022electronic}. For example, the reflectarray-based IRS demonstrated in~\cite{monroe2022electronic} is composed of nearly $10,000$ elements, while active RF chain-based arrays are yet below the 256-element mark~\cite{rodwell2019100}. By moving the burden of beamforming ``off-chip", and removing the need for any RF chains, the same IRS can be shared among multiple users, as discussed~\cite{singh2022wavefront}. IRSs can also be manipulated to create custom beam shapes~\cite{singh2022wavefront}, and if their design is realized through a metasurface-based, sub-wavelength-sized meta-atoms, the level of control is increased drastically~\cite{di2019smart, liaskos2019novel}. Such IRSs can be utilized to create anomalous reflection~\cite{liu2022review}, in which almost negligible sidelobes are produced, greatly reducing interference due to sidelobe energy and introducing new opportunities for spectrum sharing. \newrev{They can also be used to redirect beams towards different directions from where originally intended, as discussed in~\cite{shaikhanov2023remotely}}.

\new{\textit{Research challenges:}} \new{The majority of algorithms for beam design and active SLL suppression have been proposed for lower-frequency systems and, thus, do not capture challenges associated with working above 100~GHz. As we move to higher frequencies, the wavelength becomes shorter and, thus, the antenna array elements become smaller and they can be more densely packed without suffering mutual coupling~\cite{akyildiz2016realizing}. However, the size of every other component in the \gls{rf} chain (i.e., frequency multipliers, mixers, filters, and amplifiers) does not necessarily scale by the same factor. This leads to challenges in packaging of such arrays above 100~GHz. Consequently, there is a need to develop new antenna array architectures. First, one could consider adopting a fully analog antenna architecture, in which one only \gls{rf} chain is utilized to drive multiple antennas each with analog phase and/or amplitude control~\cite{sun2014mimo}.
Even in that case, there are several challenges. First, broadband phase controllers that can work across large contiguous bandwidths (e.g., tens of GHz) are still being developed. Strictly speaking, some of the benefits shown in Figure~\ref{fig:total_control} could be achieved only with the additional manipulation of the amplitude at each antenna element~\cite{vilenskiy2021quasi, balanis2016antenna}. However, the integration of an amplitude controller (ultimately, a power amplifier) per antenna element is many times not possible again not only due to packaging but also thermal issues~\cite{rodwell2019100}. 
As a solution, innovative true-time delay controllers could be utilized, for example by leveraging new materials such as graphene \newrev{specifically proposed for the terahertz band}~\cite{chen2013thz} and, accordingly, the algorithms adapted.
}

\new{Many of the challenges associated with building such antenna arrays and IRSs originate in the fact that the designs demonstrated to date leverage the so-called electronic approach to sub-THz and THz devices~\cite{akyildiz2022terahertz}, in which the limits of traditional electronic systems are pushed to operate at higher frequencies. Alternatively, again new materials could be adopted to build arrays that leverage the properties of surface plasmon polariton waves and intrinsically operate at THz frequencies~\cite{akyildiz2016realizing}. As of today, graphene-based plasmonic signal sources~\cite{jornet2014graphene,crabb2021hydrodynamic}, plasmonic phase, amplitude and frequency modulators~\cite{singh2016graphene,crabb2022amplitude}, and antennas~\cite{jornet2013graphene} have been demonstrated. Their integration in very dense arrays with element spacing much smaller than the free-space wavelength enables ultra-massive MIMO systems~\cite{singh2020design}, with unprecedented control of the beam shapes and wavefronts~\cite{singh2022wavefront}. Nevertheless, this technology is much less mature than electronic or even photonic solutions, making it a high-risk, high-reward approach. Even with IRSs, although it is much easier to scale up in size, designing high-performing, versatile analog phase shifters at such high frequencies is a challenging task~\cite{dahri2018polarization}. Among other solutions, the utilization of plasmonic materials and hybrid structures presents an interesting direction for realizing these devices~\cite{singh2020hybrid}.}

\glsreset{fss}
\textbf{Passive Mitigation Techniques.} In addition to preemptive suppression techniques at the \gls{bs}, it is possible to address interference through passive devices that do not increase the complexity of the BS itself. From the discussions in Section~\ref{subsec:sensing} and \ref{sec:interference}, it is evident that sensitive devices which \new{require} protection from unwanted interference will rarely occupy the entire spectrum of emission of the BS. For example, the presently prohibited \SIrange{109.5}{111.8}{\giga\hertz} portion of the spectrum can be specifically addressed by suppressing \gls{rfi} within this \SI{2.3}{\giga\hertz} bandwidth. In this regard, \glspl{fss} are devices that specifically address such sensitive specks of a frequency spectrum. As shown in Figure~\ref{fig:device_design_fss}, an FSS is a dense, metasurface-like configuration with resonant passive elements in an array, usually designed to intercept and interact with a bandwidth of interest~\cite{munk2005frequency, al2011wideband, li2012broadband}. These have traditionally been utilized to reduce the cross-section of radars~\cite{munk2005frequency}, and recently have been proposed to reduce unwanted \glspl{sll} and leakage in wide-band antenna arrays above 100~GHz~\cite{nissanov2021high}. In addition, these have been proposed to reduce mutual coupling effects between antenna elements in densely packed antenna arrays~\cite{zhang2019mutual}. Within the \gls{mmwave} band, \glspl{fss} have attracted attention, inter alia, to act as planar lenses with enhanced directivity, low cost, and ease of deployment~\cite{al2011wideband, li2012broadband}. With an \gls{fss}, it is possible to design both band-stop and band-pass filters that can be deployed without any additional complexity at the \gls{bs} itself, avoiding the complexity of greater control and increased power consumption requirements. The effectiveness of an \gls{fss} depends on the relative orientation of the surface with the unwanted radiation. Thus, with proper alignment towards the dominant sidelobes that may interfere with sensitive devices, an \gls{fss} designed to act as a stop-band at the sensitive portions of the spectrum can be deployed, through which the unwanted leakage can be suppressed significantly~\cite{munk2005frequency}. 

\new{\textit{Research challenges:}} \new{While simpler than the devices needed for active \gls{sll} suppression, the main challenges associated with the passive solutions relate again to the hardware design. In particular, when implemented through metasurfaces, the design of the \glspl{fss} is not governed by well-defined equations and operating theory as, for example, the design of beamforming or beam-nulling algorithms. Instead, an ``artisan" based trial-and-error methodology is required to develop the meta-atom shape and arrangement~\cite{liaskos2019novel, venkatesh2020high}. As an alternative, a library of meta-atoms designs and potential arrangements to achieve different functionalities needs to be built, which could benefit the wireless community at large. In addition, there are system level challenges associated to deploying such devices, in terms of both network and frequency planning.}





\subsection{Solutions based on \gls{rfi} Mitigation for Passive Users}

Complementary and independent of custom device design solutions, harmful \gls{rfi} can be mitigated with signal processing techniques. \gls{rfi} mitigation has been typically studied as a necessary step in passive sensing signal processing pipelines, to protect from undesired \gls{rf} emissions in the band of interest, either intentional or unintentional. As discussed in Section~\ref{sec:stakeholders}, the receivers used by several passive sensing applications (e.g., radio astronomy) have extremely high sensitivity and can pick up signals that would otherwise be considered below the noise floor~\cite{fridman2001rfi}. In the context of spectrum sharing, \gls{rfi} mitigation puts the burden of the interference management on the passive users. Therefore, while it is an essential component of the passive sensing digital signal pipeline, it represents just one of the different ingredients that need to be combined for a balanced approach to spectrum sharing. In the next paragraphs, we review general characteristics and approaches for \gls{rfi} mitigation, and discuss a few specific examples of \gls{rfi} mitigation above 100~GHz.

\begin{figure}
  \centering
  \includegraphics[width=.9\columnwidth]{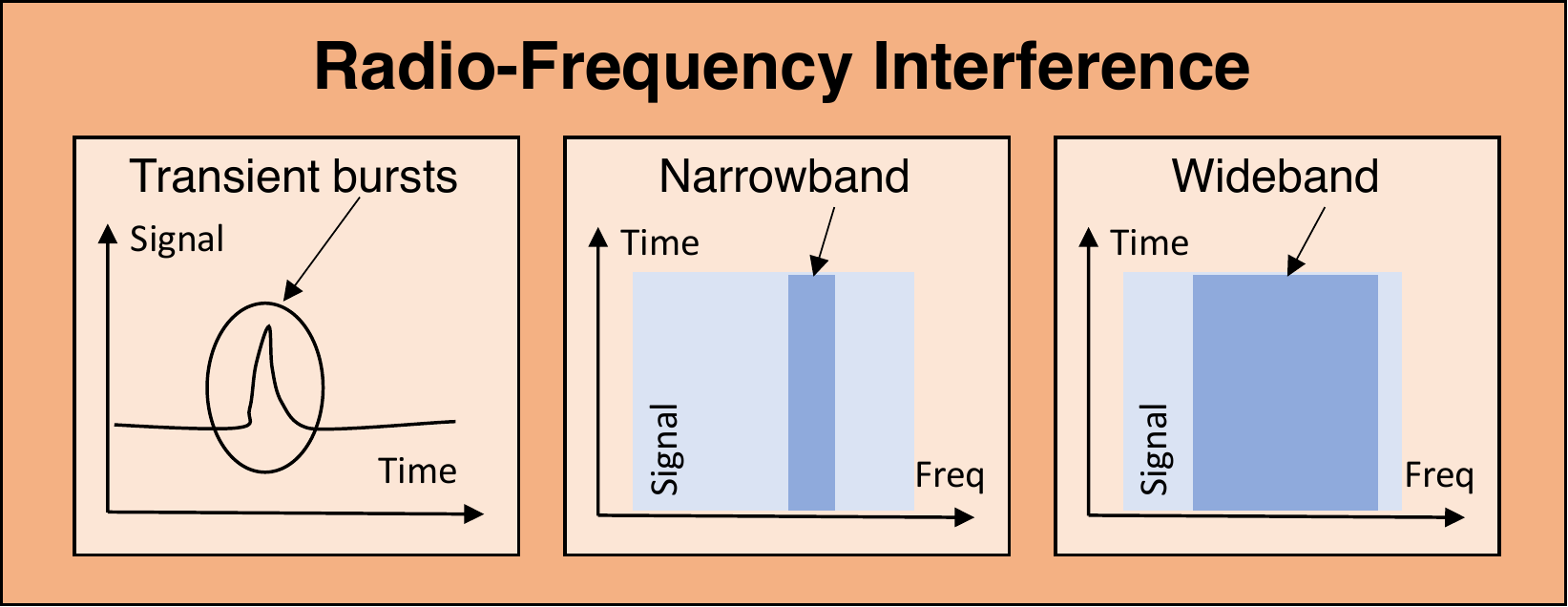}
  \caption{Classification of different sources of \gls{rfi}.}
  \label{fig:rfi}
\end{figure}

\gls{rfi} is generally categorized in three different classes (Figure~\ref{fig:rfi}), i.e., transient bursts, generated by impulse-like waveforms, and narrow-band or wide-band signals. Additionally, \gls{rfi} may be caused by sources that transmit in the band of interest, or by out-of-band emissions. The \gls{rfi} may (i) \new{introduce} clearly identifiable distortion(s) in the gain or frequency response of the received waveform; (ii) obscure the received signal, with a strong and identifiable emission in the entire observed spectrum; or (iii) introduce more subtle alterations in the received signals (e.g., if the \gls{rfi} power is close to the noise threshold), which are harder to detect and, consequently, filtered out. 

\textbf{General \gls{rfi} Mitigation Techniques.} The general approaches toward \gls{rfi} mitigation involve temporal, frequency, or spatial filtering, or rely on specific properties of the received signals. Temporal and frequency domain \gls{rfi} can be detected and blocked if confined in specific time instants or sub-bands, for example with threshold-based methods. For measurements \new{of natural phenomena}, whose statistical distribution is known (and often Gaussian), statistical methods can be used to identify and reject outliers that may correspond to \gls{rfi}~\cite{balling2012rfi}. Widely used techniques include (i) evaluating the kurtosis ratio of the received signal~\cite{bradley2014detection}, i.e., the ratio between the fourth moment and the squared second moment (variance) of a signal, which should be equal to 3 for Gaussian signals; and (ii) evaluating the third and fourth Stokes parameters of the signal, which are related to its polarization and are typically small in the absence of \gls{rfi}~\cite{balling2012rfi}. The \gls{rfi} can also be detected by analyzing signals in the spatial domain. If the direction of the source of the intended signal is known, \gls{rfi} can be detected through filtering signals with Angle of Arrival (AoA) outside a pre-defined region~\cite{fridman2001rfi}. Importantly, as discussed above, the presence of massive-MIMO arrays in the 100~GHz region provide an ability of unprecedented angular resolution and accurate AoA detection. Alternatively, highly diffused sources without \gls{rfi} exhibit a smooth behavior in the spatial and angular domain, thus \gls{rfi} can be detected and filtered by analyzing the outliers. State-of-the-art \gls{rfi} mitigation systems  often combine two or more of these techniques, to maximize the probability of detecting \gls{rfi}~\cite{piepmeier2014radio,piepmeier2017smap,johnson2016soil}.

\textbf{Use-case-specific \gls{rfi} Mitigation Techniques.} The \gls{rfi} mitigation techniques need to adapt both to the characteristics of the interferer and to the nature of the observation and equipment utilized. The result is that, oftentimes, general signal processing techniques for \gls{rfi} mitigation need to be adapted to the specific use case~\cite{querol2015assessment,fridman2001rfi,johnson2016soil,PECK201354}. Signal processing for \gls{rfi} mitigation can be run offline~\cite{Zhang_2020} or online~\cite{van2016towards}. While this may be a concern for the spectrum above 100~GHz, given the large bandwidth at stake, the research and development of hardware-accelerated signal processing pipelines has led to multi-carrier filter banks capable of processing tens of GHz of bandwidth~\cite{vidunethIRMMWTHz}. Similarly, mixed online and offline approaches are emerging to reap the benefits of both, e.g., for the \gls{rfi} mitigation system of the next-generation Very Large Array (ngVLA) radio telescope~\cite{selina2020rfi}. This enables both real-time adaptability to the source and more complex signal processing offline, with applicability to very large bandwidths. While the ngVLA is expected to work in the 1.2-116~GHz band, the hardware and software techniques developed for its \gls{rfi} mitigation could be re-used in above-100-GHz systems as well.

\new{\textit{Research challenges:}} 
\new{So far, most of the research on \gls{rfi} mitigation has focused on sensing bands below 24~GHz~\cite{levine2016rfi}, which were considered the most exposed to undesired emissions from commercial radios. This represents a potential issue for current systems operating above 100~GHz, as they are generally not equipped with \gls{rfi} mitigation pipelines,  which reduces the ability for sharing and coexisting with active systems. Future sensing instruments can instead be equipped with \gls{rfi} mitigation mechanisms, whose design, however, is challenged by a number of elements. As previously mentioned, the large bandwidth complicates the design of real-time signal processing pipelines for \gls{rfi} mitigation. In addition, the design of \gls{rf} circuits is made more complex by the high frequency and large bandwidth requirements, which may introduce greater interference from out-of-band emissions. However, the limited multipath (or absence of it) \newrev{typical of propagation above 100 GHz}~\cite{han2021hybrid} makes spatial filtering extremely powerful, as \gls{rfi} sources exhibit limited or no scattering and the number of directions to target for nulls become smaller.}

\subsection{Solutions based on Waveform Engineering and Digital Signal Processing for Active Users}

It is further possible to reduce RFI through waveform designs where the communication and sensing signals are made orthogonal to each other, or are otherwise engineered to not register as interfere with each other. 

\textbf{Spread Spectrum Techniques.} A very large contiguous bandwidth can be leveraged to create spread spectrum waveforms, in which the communication signal can be artificially lowered to below the noise floor. Such methods have been utilized in \gls{dsss} to create efficient spectrum sharing, such as in \gls{cdma}, where a specific communication signal is multiplied by a spreading key, to artificially spread a narrow-band signal over a very large bandwidth~\cite{zigangirov2004theory}. The receiver utilizes the same key to recover the signal, and all other signals within the same sub-space are filtered as noise. In this setup, the larger the pseudo-random key generating sequence, the greater the number of orthogonal signal sets that can be created. In addition, by combining multiple communication bands together, frequency hopping, similar to as in Bluetooth transmission, can be leveraged to create multiple windows of signal transmission that fall outside the critical bandwidth portions reserved for sensing. Ultimately however, such spreading techniques compromise the very high data rate requirement for communication standards, and would require a bandwidth which proportionally increases with the length of the sequence, as compared to that presented in Figure~\ref{fig:datarates_mimo}.

\begin{figure}[t]
  \centering
  \includegraphics[width=\columnwidth]{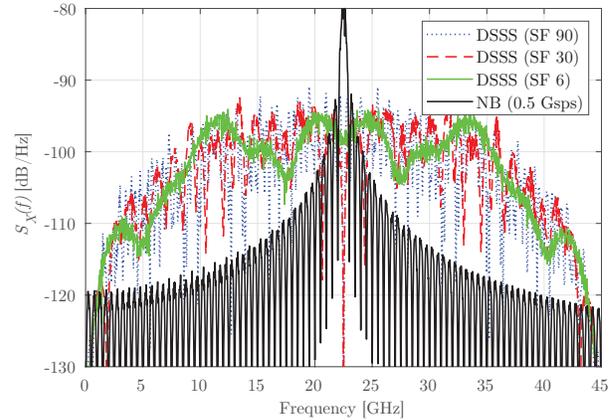}
  \caption{Power spectral density of a \gls{dsss} signal with different SF and a narrowband signal, both modulated with 4-QAM. Figure adapted from~\cite{bosso2021ultrabroadband}.}
    \label{fig:PSD_all}
\end{figure}

\new{\textit{Research challenges:}} \new{As of today, the main bottleneck in \gls{dsss}-based coexistence techniques above 100~GHz is the real-time generation of ultra-broadband spreading sequences. For example, in~\cite{bosso2021ultrabroadband}, we built and experimentally demonstrated a \gls{dsss} system for sharing between active and passive systems above 100~GHz. The framework can successfully generate sequences at 130~GHz with a 20~GHz spread bandwidth. Figure~\ref{fig:PSD_all} shows the \gls{psd} over frequency for a narrowband signal and \gls{dsss} signals with three different spreading factors, all for a 4-QAM signal. An increase of the spreading factor leads to more uniform upper envelopes for the \gls{psd}, with peaks at about 15 dB below the peak of the narrowband signal, at the cost of achievable data rate. 
This effect, which is popular in tactical communications contexts to decrease the signal interception or eavesdropping probability, can also lead to received signals at the satellite sensor which are below the \gls{rfi} threshold, as further discussed in~\cite{bosso2021ultrabroadband}. Indeed, a reduction of 15 dB in the received signal at the passive system can increase the parameter space in which active systems can coexist without exceeding the \gls{rfi} threshold, e.g., when compared to the configurations in Figure~\ref{fig:interference}.
However, to realize this, we used a state-of-the-art arbitrary-waveform generator with data-converters operating at 92~Gigasamples-per-second at the transmitter and 160~Gigasamples-per-second at the receiver, with power requirements, size and cost only suitable for lab testing. To overcome the bottleneck introduced by the high-speed data-converters, one solution is to design fully analog \gls{dsss} chips that operate directly with analog I/Q samples in baseband or at an IF frequency. The design of such chips should be accompanied by the exploration of low-complexity real-time implementable orthogonal sequence generation algorithms.}

\textbf{Frequency-Selective Waveforms.} Transmitters which can support very high \gls{papr} can be utilized to actively remove compromised portions of the signal space from waveforms. An example where such a setup could potentially be implemented without significantly reducing the communication capacity is in \gls{ofdm}~\cite{benvenuto2021algorithms,banelli2014modulation,berardinelli2014potential}. 
%
\new{In OFDM, groups of subcarriers are organized into resource blocks, and power allocation can be performed on a resource block level.}
The judicious allocation of resource blocks determines the peak capacity rate. In such a setup, it may then be possible to purposely \new{identify} sub-carriers within the sensitive regions and minimize the effective power distributed to them. 

\new{\textit{Research challenges:}} \new{In alignment with many of the previous challenges, the bottleneck is posed again by the hardware and, particularly in this case, the power amplifiers. At all frequencies but particularly at frequencies above 100~GHz, power amplifiers are very sensitive to power fluctuations and, thus, carrier nulling in \gls{ofdm} is many times not an option. Besides, waiting on the development of more robust amplifiers, there are other possible solutions. For example, a common waveform found in radar systems, also those at frequencies above 100~GHz, relies on continuous frequency modulation or chirp signals. Recently, our group has demonstrated the ability to utilize chirps in Chirp Spread Spectrum (CSS) to both transmit information even in extreme frequency-selective channels (such as those over an absorption line at THz frequencies)~\cite{sen2020ultra} and to build a joint communication and atmospheric radar system~\cite{aliaga2022joint}. Differently, one could design non-continuous frequency chirps that skip certain frequency bands to minimize the radiation in forbidden bands.}

\textbf{Orthogonal Waveforms.} Another technique, only valid for the coexistence of active sensing (e.g., radar) and communication users, relies on utilizing waveforms generated in the context of orthogonal basis sets, where every signal is independent and orthogonal to the other. One such example is the Gram-Schmidt orthogonalization~\cite{grami2015introduction}. In this setup, a set of $M$ finite energy symbols is mapped to a set of $N$ orthogonal symbols as shown in Eq.~\eqref{eqref:ortho}: 
\begin{equation}
    \label{eqref:ortho}
    {s_{i}(t)}, i = 1,2,...M \mapsto {\phi_{i}(t)}, i = 1,2,...N; N \leq M.
\end{equation}
\newrev{Here, the orthogonal signals $\phi_{i}(t)$ are created from $s_{i}(t)$ through a filter bank, where the components of the previous orthonormal signals are removed from the 
concerned signal.} Gram Schmidt orthogonalization is actively used in most MIMO applications involving coherent detection \new{by utilizing QR decomposition~\cite{luethi2008gram}}. \new{Without loss of generality, assume that the first orthogonal signal, $\phi_{1}(t)$, is actively selected for sensing applications. It is then possible to eliminate any interfering portions from the remaining symbols, $\{\phi_{2,}(t),\phi_{N}(t)\}$, used for communications. This procedure requires some level of interaction between the active sensing and communications systems to coordinate on the nature of the sensing signal $\phi_{1}(t)$.} 

\newrev{It is important to note that the exclusion of an orthonormal basis function does not simply remove one symbol from the possible symbol set of transmission.  In fact, this makes the symbol set incomplete, reducing it from $M = 2^N$ to $M-1 = 2^N - 1$, thus making the representation of N bits impossible. Thus, with the capability of representing only $N - 1$ bits, only $2^{(N-1)} = M/2$ symbols are required.}

\newrev{Here, we observe that this setup can still be utilized when required to preserve multiple sensing frequencies since there are several redundant message symbols that can be utilized and thus could be utilized without losing any further capability. Ultimately however, the principles of utilizing the Gram Schmidt orthonormalization require careful consideration of the richness of the basis set, the complexity of the modulation scheme, and the modifications that must be performed with regard to ensuring the key system parameter such as the bit error rate, the data rate, and the Euclidian separation of the constellation points can be determined accurately and predictably.}


\new{\textit{Research challenges:}} \new{This type of solution is only applicable to future active and passive coexisting systems, \newrev{as it requires coordination.} As mentioned in Section~\ref{sec:burdgen}, this is the type of ``burden sharing'' expected from Resolution 731. As with the other waveforms-based solutions, the computational complexity of the modulation/demodulation algorithms, the requirements on the data-converters, and the potential implementation of the solution in the analog domain are aspects to be taken into account.} 
\newrev{In addition, the modulations utilized in high data rate communications often require precise inputs and constellation requirements, which makes the omission of a basis function particularly challenging. Certain parameters such as a uniform constellation, the efficiency of a message signal, and a possible non-completeness of the symbol set could result in non-linear distortions as well as reduced throughput, as discussed above.}







\subsection{MAC Layer and Networking-based Sharing}

Dynamic spectrum access can be enabled at the \gls{mac} layer and above through time and spatial sharing. In particular, it is possible to envision sharing mechanisms with different levels of integration among the technologies used by interested stakeholders of the above 100~GHz band, from a fully integrated solution to a \gls{cbrs}-like model.

\textbf{Shared Active / Passive Medium Access Schemes.} One option is represented by \gls{tdma} between two or more technologies, with active and passive users sharing different time resources. For example, next-generation wireless network \gls{mac} designs may include some time slots that can be blanked for sensing~\cite{shi2016coordinated}, either active or passive. To fully exploit this flexibility, the sensing stack could be integrated in the communications stack, e.g., satellites used for commercial downlink communications could be also equipped with passive sensing systems activated on a time-sharing basis. Alternatively, passive sensing systems could be equipped with an additional sensing loop that detects when the communication link is not used and collect valid observations. 

\begin{figure}
    \centering

\begin{subfigure}[b]{\columnwidth}
    \centering
    \includegraphics[width=.9\columnwidth]{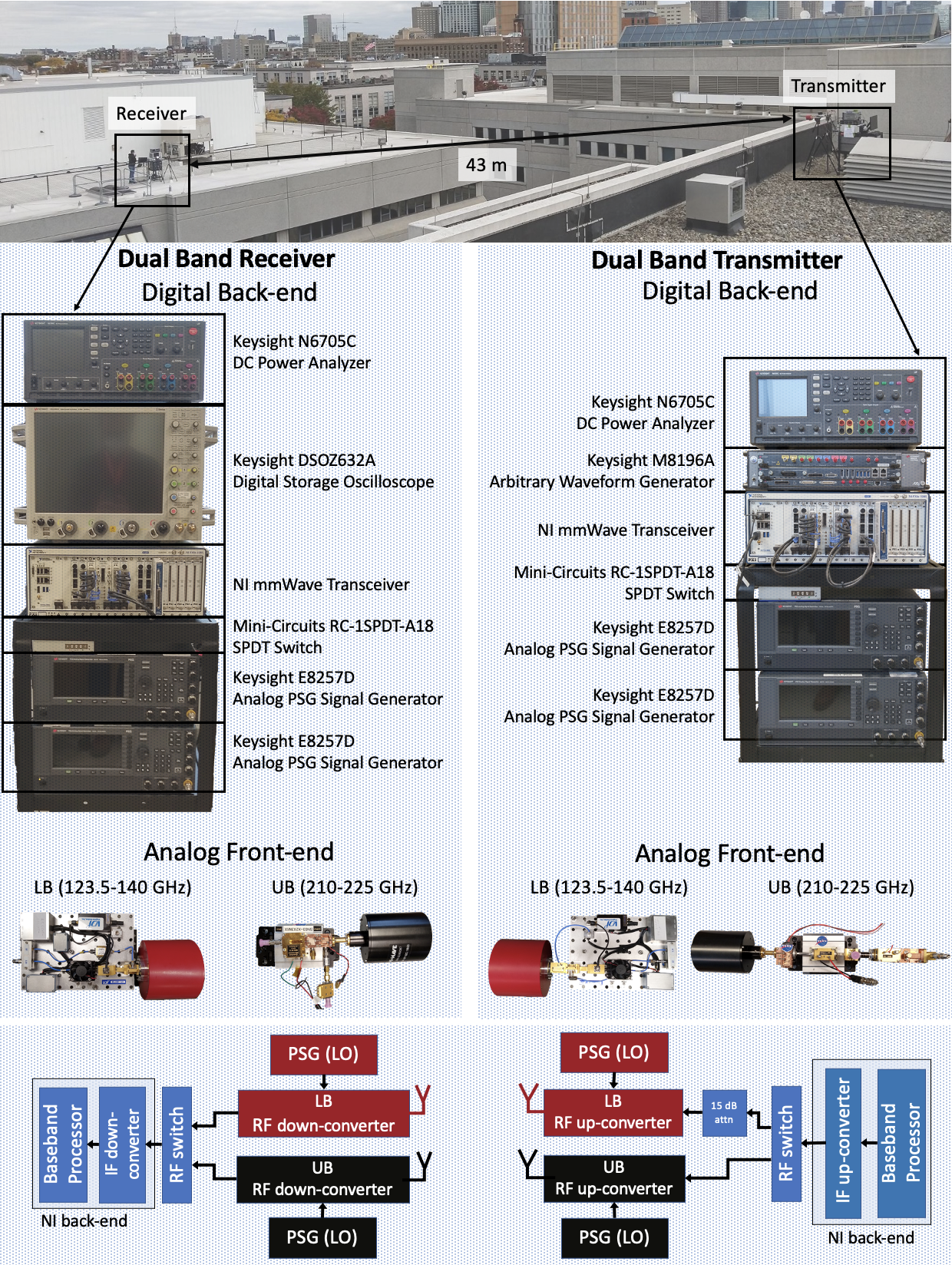}

    \caption{Prototype schematics of a dual-band prototype for dynamic spectrum access above 100~GHz, from~\cite{polese2021dynamic}.}
    \label{fig:prototype}
\end{subfigure}
\ifexttikz
    \tikzsetnextfilename{switching_distribution}
\fi
\begin{subfigure}[b]{\columnwidth}
        \centering
        \setlength\fwidth{.8\columnwidth}
        \setlength\fheight{.4\textwidth}
        \input{Figures/switching_distribution.tex}
        \caption{\gls{cdf} of the duration of a procedure to switch between bands, with a centralized (ccordinated) or distributed (indipendent) approach, from~\cite{polese2021dynamic}.}
        \label{fig:switch}
\end{subfigure}

    \caption{Prototype and performance of the first dynamic spectrum access system above 100~GHz, from~\cite{polese2021dynamic}.}
    \label{fig:dsa-100ghz}
\end{figure}

\new{\textit{Research challenges:}} \new{It is challenging to design, deploy, and implement these approaches in practice. Indeed,} they require a tight synchronization between different signal processing chains and RF circuits, which generally have a transient behavior and cannot switch on and off instantaneously\new{~\cite{peng2020channel}}. \new{Therefore, even thought this is a MAC-level strategy, it requires tight coordination and innovations in physical layer and at the device level, \newrev{especially as the duration of channel access decreases with the increase of the data rate}.} Additionally, while frequent time-switching is well tolerated by packet-based communications systems~\cite{kihero2018inter,patriciello20185g}, it may not be compatible with the sensing of continuous sources of electromagnetic signals and/or long integration times. \new{Finally, these sharing solutions are dependent on protocols that would need to be developed, tested, and standardized in the relevant forums.}

\textbf{Multi-band Systems.} It is also possible to design time sharing mechanisms with a granularity in the order of minutes. \new{While these do not enable the same level of multiplexing and spectrum utilization as MAC-based strategies, they require limited or no coordination between the users of the spectrum.}
Notably, there are approaches in which only one of the two systems (i.e., communications or sensing) can change to adapt to the other. For example, in~\cite{polese2021dynamic} we demonstrated the feasibility of passive/active spectrum sharing through dynamic spectrum access, with the communications stack dynamically adapting to the presence of passive sensing satellites over the deployment area. The system (shown in Figure~\ref{fig:prototype}) is capable of tracking satellite mobility (e.g., the NASA Aura satellite~\cite{waters2006mls}) and predict when a passive-sensing station starts orbiting over the deployment area of the communication link. When this happens, the systems switches from the band that may interfere with the satellite to another band, enabling time-sharing of resources that would otherwise be unused, while introducing no harmful interference to the sensing system. This process happens with a granularity of minutes, i.e., it does not need a tight integration between the sensing and communications stacks. Nonetheless, it is important to perform the switch between bands in a timely fashion, so as to avoid downtime for the communication link. Figure~\ref{fig:switch} compares the \glspl{cdf} of the duration of switch events with a coordinated (or centralized) approach and an independent (or distributed) strategy. Even though this is a prototype, the system achieves low switching times (i.e., in the order of ms) with the centralized switching, showing that dynamic spectrum access is feasible even in the spectrum above 100~GHz.

\new{\textit{Research challenges:}} \new{There are several system-level challenges that need to be addressed to make these schemes scalable and practical. As for the MAC-based approaches, there is a need for protocols and standardized strategies. In this sense, the transition toward more open and programmable systems in cellular networks~\cite{polese2022understanding} will help embedding dynamic control of the cellular stack and provide practical primitives to implement this kind of spectrum sharing. Besides this, other system-level challenges are related to security, scalability, and resource scheduling, in particular considering the directional nature of communications above 100 GHz and the number of users that will need to be supported in 6G systems.}

\textbf{Coordination-based Sharing.} From a broader perspective, dynamic spectrum access can be generically extended to support the coexistence of multiple services, for example through a shared, dynamic spectrum marketplace (which could be managed by the ITU or by local regulators) in which licenses can be leased dynamically, with a sub-second latency. As previously mentioned, the \gls{cbrs} band proceedings have shown that coordination and preemption among different active users is feasible on a dynamic basis. In~\cite{zylinium2019sc2report,baxley2019team}, the authors demonstrate a dynamic system for active/passive users sharing as part of one of the DARPA Spectrum Collaboration Challenge (SC2). The availability of fast connectivity and data bases is a key step toward the feasibility of distributed spectrum access systems that can control spectrum usage with granularity of tens of kilohertz in bandwidth and 1~ms in time (typical values for 5G systems)~\cite{zylinium-pawr}. 
Coordination-based solutions are also mentioned and evaluated in the National Research Council report on how spectrum can be used by scientific users in the 21st century~\cite{nrc2010spectrum}. The authors envision a scenario with multiple satellites for Earth exploration, each with a beam covering an area on Earth with a diameter of 30~km. The active users on the ground could coordinate with the sensing satellites through a shared database and blank the resources required for sensing on an ad hoc basis. For a constellation of 30 satellites, with the aforementioned characteristics, the authors of~\cite{nrc2010spectrum} estimate that active users can transmit in the band of interest of the passive sensing system for 99.7\% of the time \textit{without causing any interference}. Similar results are presented in~\cite{eichen2019real}. This shows that spectrum sharing between active and passive users is possible, with the proper combination of policies and technologies.

\new{\textit{Research challenges:}} \new{As for the previous higher-layer techniques, most challenges arise at the system level, in particular when considering the number of active and passive users that require coordination, the size of the database, and the need for real-time adaptation. A promising factor is that there is a limited number of passive sensing facilities, whose position and frequency occupation is often know or predictable, and by the fact that terrestrial sensing stations (e.g., for radio astronomy) are in remote locations with limited human presence. Nonetheless, any solution that has the potential to be deployed in practical scenarios needs to be tested for robustness, scalability, and security, without harming passive users in the process, while guaranteeing the data rates and quality of services levels expected for above-100-GHz communications systems. 
Finally, sensing- and beaconing-based coordination (e.g., as \gls{cbrs}) are challenged by the directional nature of communications above 100 GHz, which could introduce deafness to some transmissions or beacons.}



\section{Conclusion}
\label{sec:conclusion}

The future of technologies and policies in the spectrum above 100~GHz is being shaped by the development of new use cases and capabilities in the communications and active and passive sensing domains. In this paper, we have provided the first comprehensive overview of stakeholders, policies, and technologies for spectrum sharing in the band above 100~GHz. First, we reviewed the relevance of this portion of the spectrum for sensing applications in both the active and passive sensing domains, and for next-generation communication networks. We discussed the need for more bandwidth for improved performance in both sensing and communication applications, which motivates spectrum sharing approaches, and reviewed the relevant spectrum policies in this domain. We then analyzed the impact that active terrestrial users may have on passive sensing satellites, showing how spectrum sharing techniques need to protect and account for the presence of passive incumbents. Finally, we discussed possible approaches to spectrum sharing above 100~GHz, considering devices, signal processing, and higher-level coordination.

\textbf{Outlook --- Spectrum Policies for the Next Decades} The paper clearly shows that there exists need, opportunities, and technologies to enable a shared use of the spectrum above 100~GHz. In this sense, the sensing and communication communities have the opportunity to work together to evolve spectrum policies and technologies. Efforts in this direction can lead to:
\begin{itemize}
    \item enabling experimental research on spectrum sharing between active and passive users, fully realizing the provisions in the ITU Resolution 731~\cite{itu2019res731}, and jumpstarting the development of technologies for safe spectrum sharing above 100~GHz. In this sense, the development of \glspl{nrdz} will be critical for the study of active/passive coexistence. The U.S. National Science Foundation \gls{nrdz} program seeks to establish large-scale testbeds in contexts where \gls{rf} transmissions outside current regulations will not harm normal receivers. Through this program, we are developing \gls{nrdz} capabilities in Colosseum, an open-access and publicly available hardware-in-the-loop wireless network emulator~\cite{bonati2021colosseum}. Colosseum will be extended to support large-bandwidth, highly directional transmissions (such as those of interest in the spectrum above 100 GHz), and is naturally positioned for spectrum sharing studies without \textit{any} risk of harming legitimate users of the spectrum thanks to its emulation capabilities;
    \item identifying opportunities for a sharing-aware approach for the development of future scientific sensing equipment and communication networks. For example, active engagement across different communities could lead to the establishment of sharing procedures similar to that of the \gls{cbrs} band, with benefit for all parties involved;
    \item using an evidence-driven, physics-based approach to develop spectrum policies that involve sharing in the RR5.340 bands, without harming passive incumbents. 
\end{itemize}


\ifCLASSOPTIONcaptionsoff
  \newpage
\fi


\bibliographystyle{IEEEtran}
\bibliography{./Bibliography/bibliography.bib}


\begin{IEEEbiography}[{\includegraphics[width=1in,height=1.25in,clip,keepaspectratio]{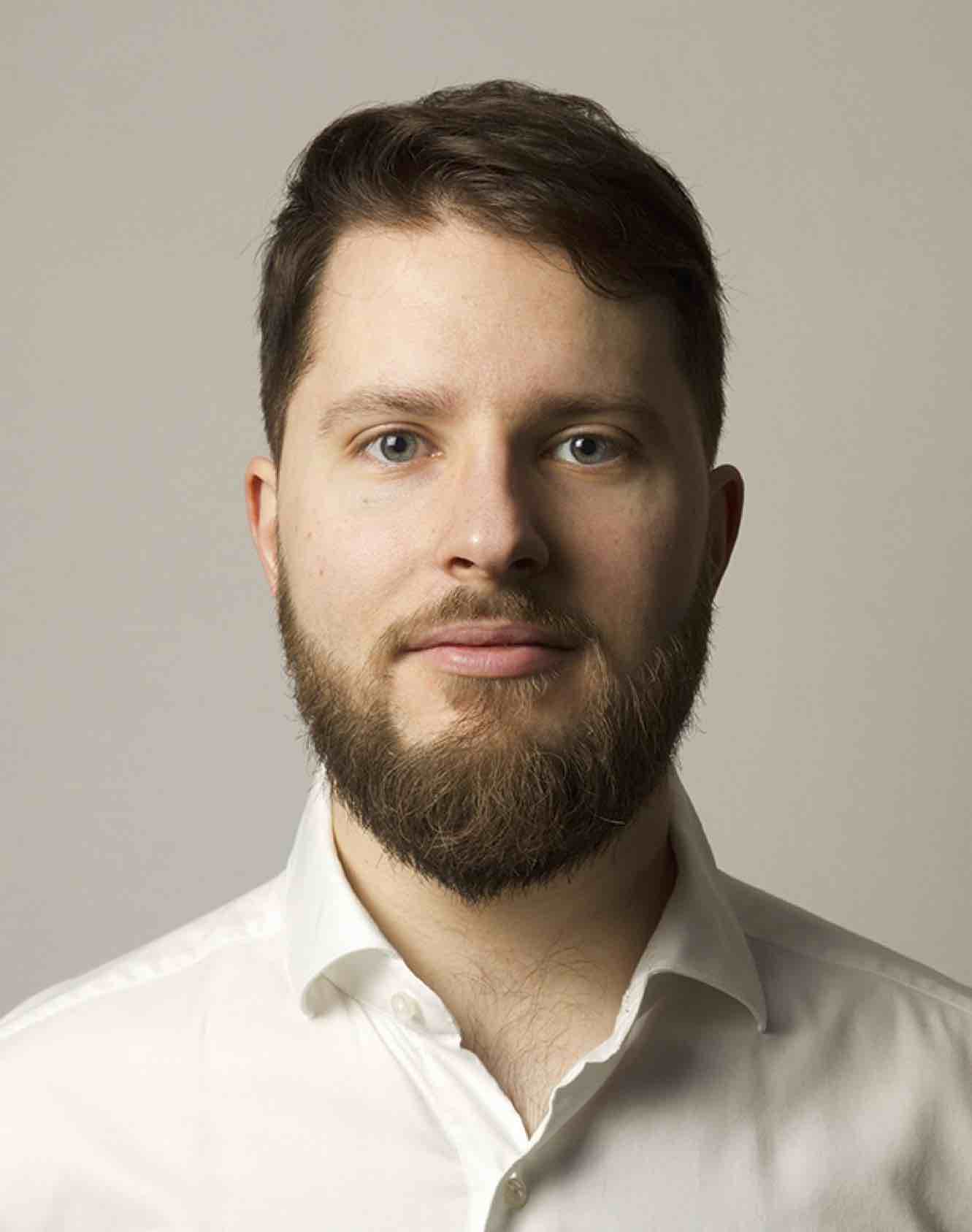}}]{Michele Polese} (S'17, M'20) is a Principal Research Scientist at Northeastern University, Boston, since March 2020. He received his Ph.D. at the Department of Information Engineering of the University of Padova in 2020 under the supervision of with Michele Zorzi. He also was an adjunct professor and postdoctoral researcher in 2019/2020 at the University of Padova, and a part-time lecturer in Fall 2020 and 2021 at Northeastern University. During his Ph.D., he visited New York University (NYU), AT\&T Labs in Bedminster, NJ, and Northeastern University. He collaborated with several academic and industrial research partners, including Intel, InterDigital, NYU, AT\&T Labs, University of Aalborg, King's College and NIST.
He was awarded with several best paper awards, and is serving as TPC co-chair for WNS3 2021-2022 and as an Associate Technical Editor for the IEEE Communications Magazine. His research interests are in the analysis and development of protocols and architectures for future generations of cellular networks (5G and beyond), in particular for millimeter-wave communication, and in the performance evaluation of complex networks.
 \end{IEEEbiography}
 
 \begin{IEEEbiography}[{\includegraphics[width=1in,height=1.25in,clip,keepaspectratio]{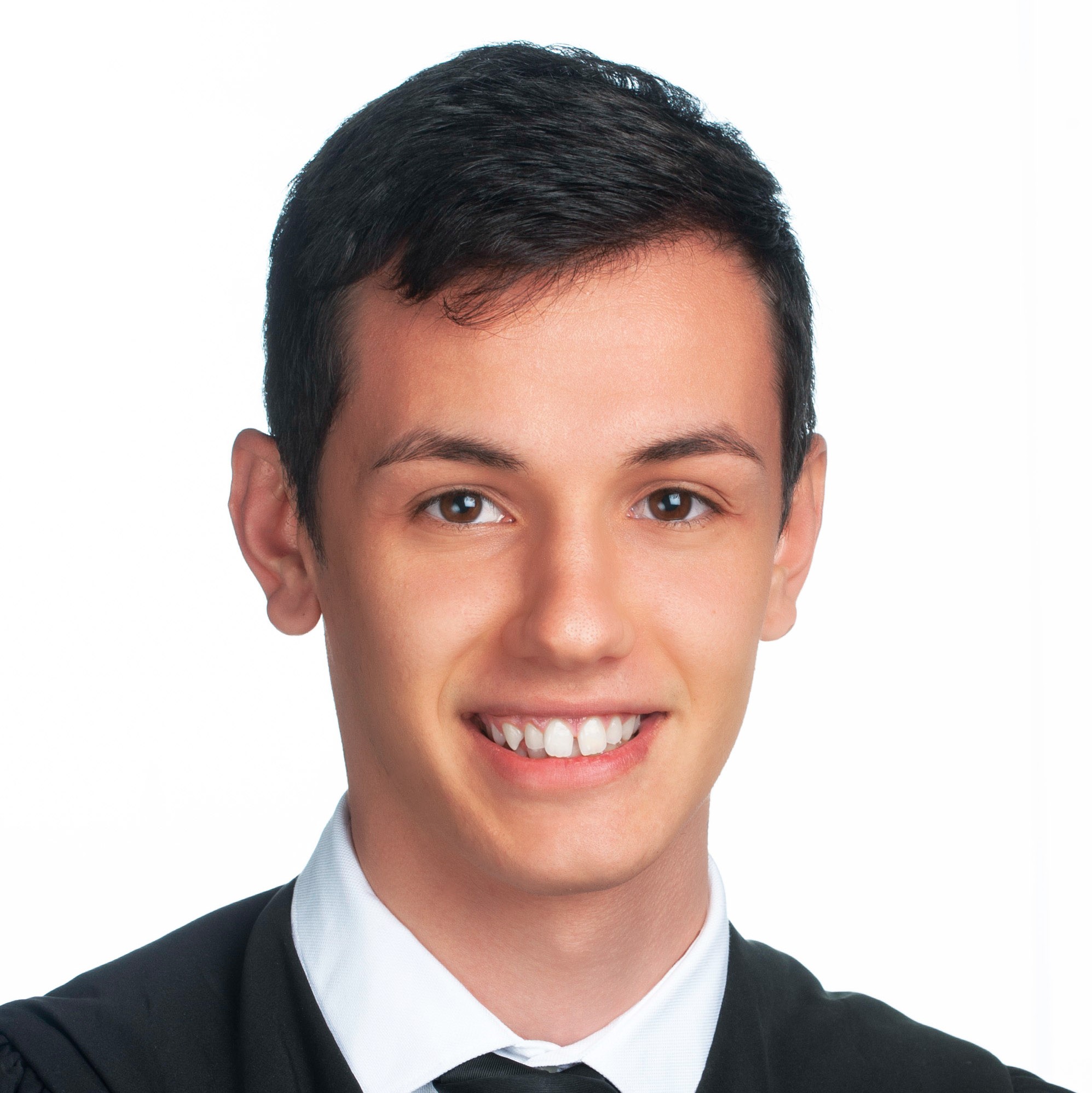}}]{Xavier Cantos-Roman} (Student Member, IEEE) received the B.S. in Telecommunications Technologies and Services Engineering from the Universitat Politècnica de Catalunya, Barcelona, Spain, in 2019. He is a member of the Ultra-broadband Nanonetworking Laboratory at Northeastern University, Boston, MA, where he is currently a Ph.D. candidate under the guidance of Dr. Josep M. Jornet. His research interests include the modeling and simulation of THz plasmonic devices, the study of mmWave and THz signals propagation and the design and experimental demonstration of joint communication and sensing systems at sub-THz frequencies.
 \end{IEEEbiography}
 
\begin{IEEEbiography}[{\includegraphics[width=1in,height=1.25in,clip,keepaspectratio]{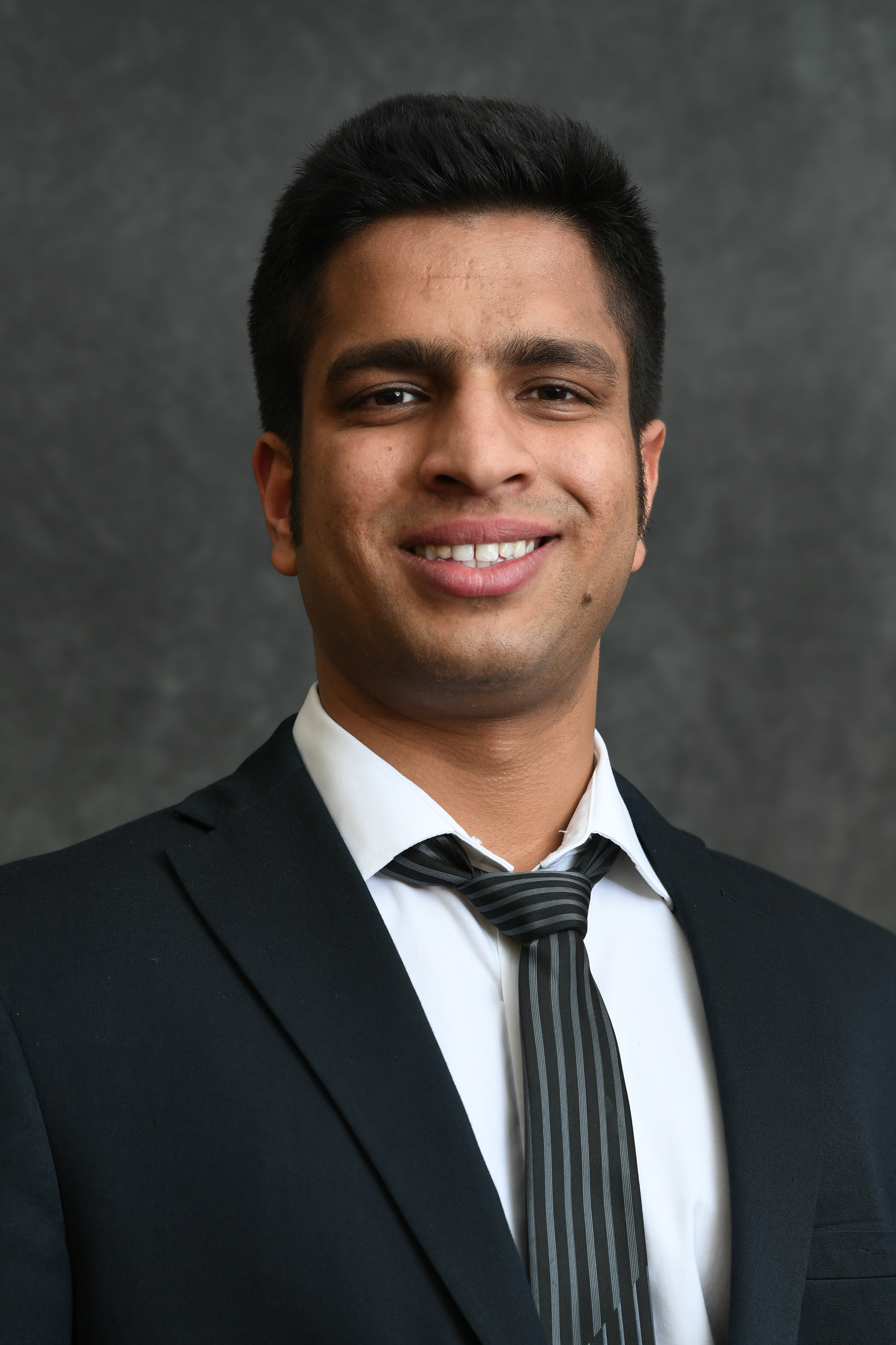}}]{Arjun Singh} (Member, IEEE)  received the B.S. summa cum laude, and M.S. in Electrical Engineering from the University at Buffalo, The State University of New York, NY, USA, in 2016 and 2018, respectively. He obtained his Ph.D. in Electrical Engineering from Northeastern University, Boston, USA, under the guidance of Dr. Josep Jornet, as a member of the Ultra-Broadband Nanonetworking Laboratory, in December, 2021.  Since 2022, he is an Assistant Professor in the Department of Engineering at the State University of New York Polytechnic Institute, Utica, NY. His research interests include realizing Terahertz-band wireless communications, dynamic spectrum sharing, space networks, wavefront engineering, graphene-plasmonics, and intelligent reflecting surfaces. In these areas, he has coauthored several publications in leading journals, as well as 1 US patent. He is also serving as the media chair for the IEEE RCC Special Interest Group on Terahertz Communications and as a reviewer for reputed journals including IEEE communications magazine. 
\end{IEEEbiography}

\vspace{2cm}

\begin{IEEEbiography}[{\includegraphics[width=1in,height=1.25in,clip,keepaspectratio]{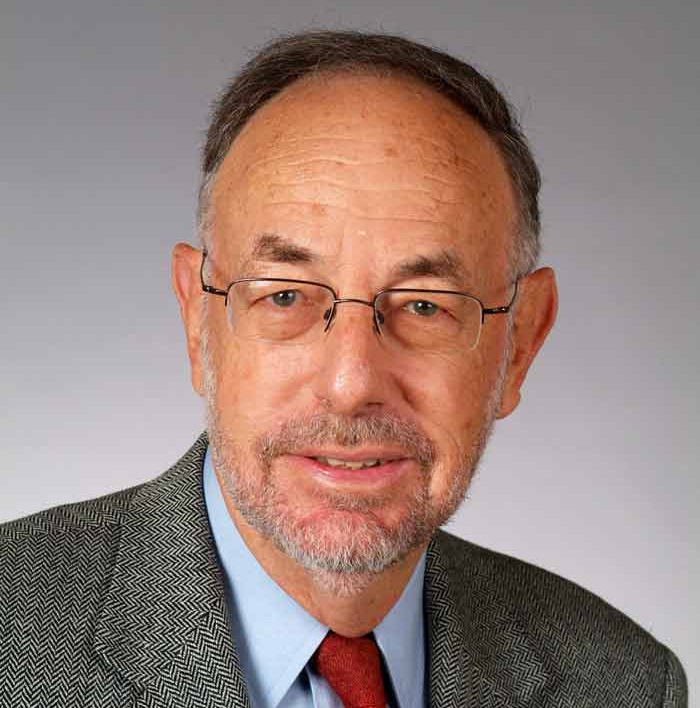}}]{Michael Marcus} (F'04) is Director of Marcus Spectrum Solutions, Cabin John, Maryland and adjunct professor at Northeastern University's Department of Electrical and Computer Engineering, Boston, MA. He retired from the Federal Communications Commission in 2004 after nearly 25 years in senior spectrum policy positions.  While at FCC, he proposed and directed the policy developments that resulted in the bands used by Wi-Fi, Bluetooth, and licensed and unlicensed millimeter wave systems above 59 GHz. He was an exchange visitor from FCC to the Japanese spectrum regulator (now MIC) and has been a consultant to the European Commission and the Singapore regulator (now IMDA). During 2012-13 he was chair of the IEEE-USA Committee on Communication Policy and is now its vice chair for spectrum policy.  In 2013, he was awarded the IEEE ComSoc Award for Public Service in the Field of Telecommunications "for pioneering spectrum policy initiatives that created modern unlicensed spectrum bands for applications that have changed our world".  He received S.B. and Sc.D. degrees in electrical engineering from MIT.
\end{IEEEbiography}

\begin{IEEEbiography}[{\includegraphics[width=1in,height=1.25in,clip,keepaspectratio]{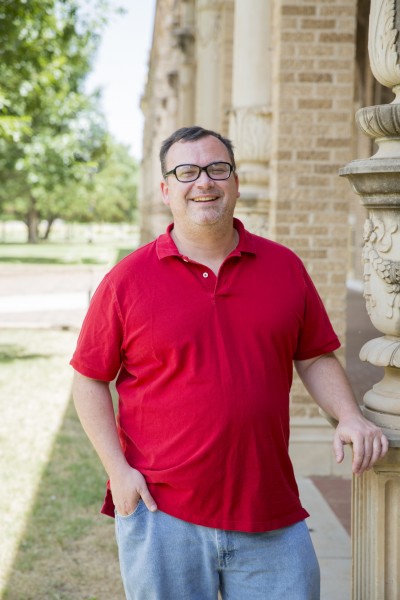}}]{Thomas J. Maccarone}  was born on August 26, 1974, in Haverhill, Massachusetts in the USA.  He received a BS in physics from the California Institute of Technology in 1996, MS and MPhil degrees in 199 and a PhD in 2001 from Yale University in astronomy.  After postdoctoral fellowships at the Scuola Internazionale Superiore di Studi Avanzati in Trieste, Italy and the University of Amsterdam, he was a lecturer then reader in the School of Physics of Astronomy at the University of Southampton from 2005-2012.  In 2013, he moved to Texas Tech University as an associate professor in the Department of Physics to help start the university's astrophysics research program, and now is Presidential Excellence in Research Professor in its Department of Physics \& Astronomy.  His main area of research is understanding the physics of accretion and radio jet production in X-rays binaries as well as the processes of binary formation and evolution for these systems.  He has been heavily involved in efforts to plan the science cases for a wide range of new observational facilities.  He serves on advisory panels to the National Radio Astronomy Observatory and the Executive Committee of the High Energy Astrophysics Division of the American Astronomical Society.
\end{IEEEbiography}

\begin{IEEEbiography}[{\includegraphics[width=1in,height=1.25in,clip,keepaspectratio]{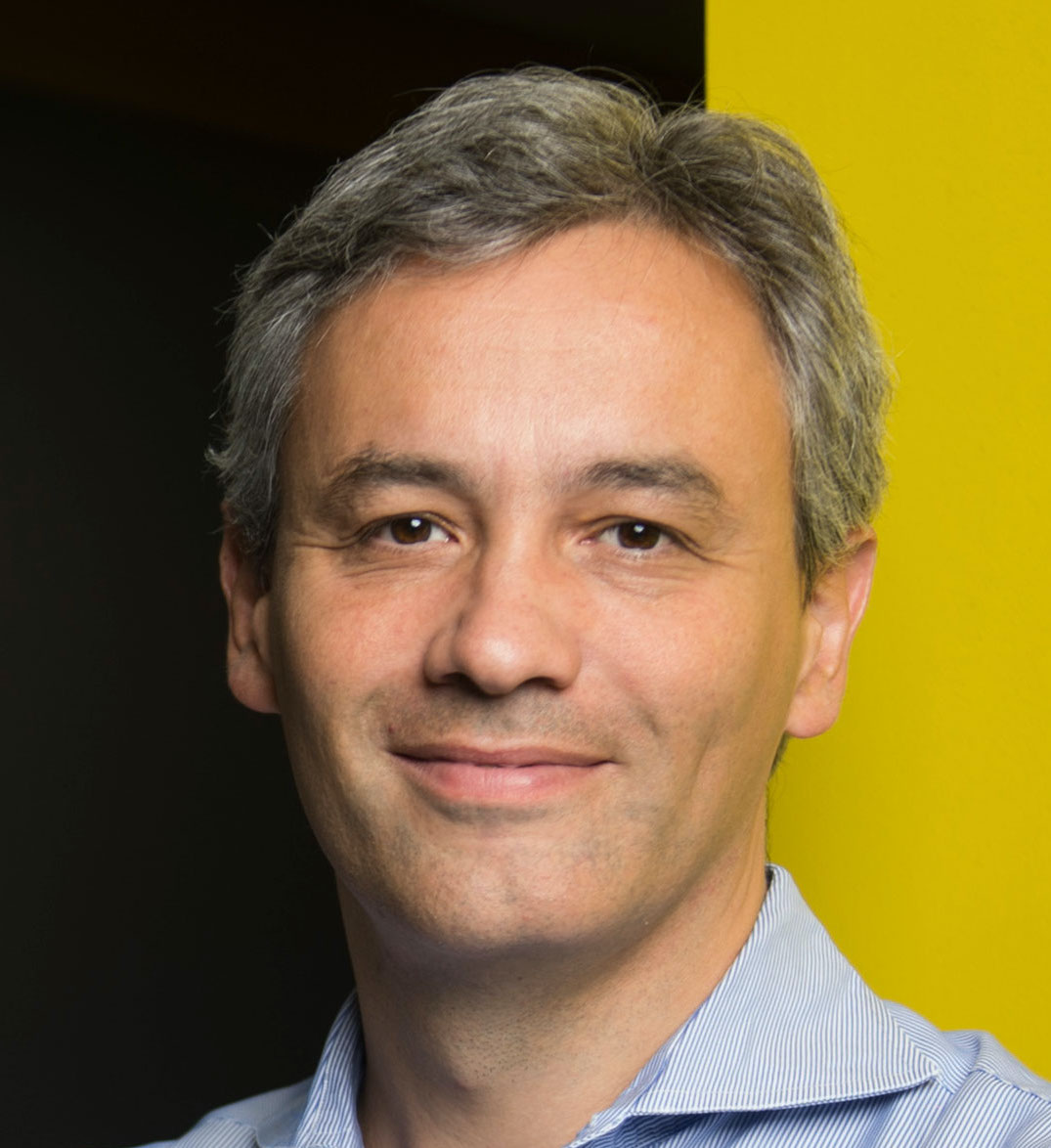}}]{Tommaso Melodia} (F'18) is the William Lincoln Smith Chair Professor with the Department of Electrical and Computer Engineering at Northeastern University, Boston. He is also the Founding Director of the Institute for the Wireless Internet of Things. He received his Ph.D. in Electrical and Computer Engineering from the Georgia Institute of Technology in 2007. He is a recipient of the NSF CAREER award. Prof. Melodia has served as Associate Editor of IEEE Transactions on Wireless Communications, IEEE Transactions on Mobile Computing, Elsevier Computer Networks, among others. He has served as Technical Program Committee Chair for IEEE Infocom 2018, General Chair for IEEE SECON 2019, ACM Nanocom 2019, and ACM WUWnet 2014. Prof. Melodia is the Director of Research for the Platforms for Advanced Wireless Research (PAWR) Project Office, a \$100M public-private partnership to establish 4 city-scale platforms for wireless research to advance the US wireless ecosystem in years to come. Prof. Melodia's research on Internet-of-Things and wireless networked systems has been funded by the National Science Foundation, the Air Force Research Laboratory the Office of Naval Research, DARPA, and the Army Research Laboratory. Prof. Melodia is a Fellow of the IEEE and a Senior Member of the ACM.
\end{IEEEbiography}

\begin{IEEEbiography}[{\includegraphics[width=1in,height=1.25in,clip,keepaspectratio]{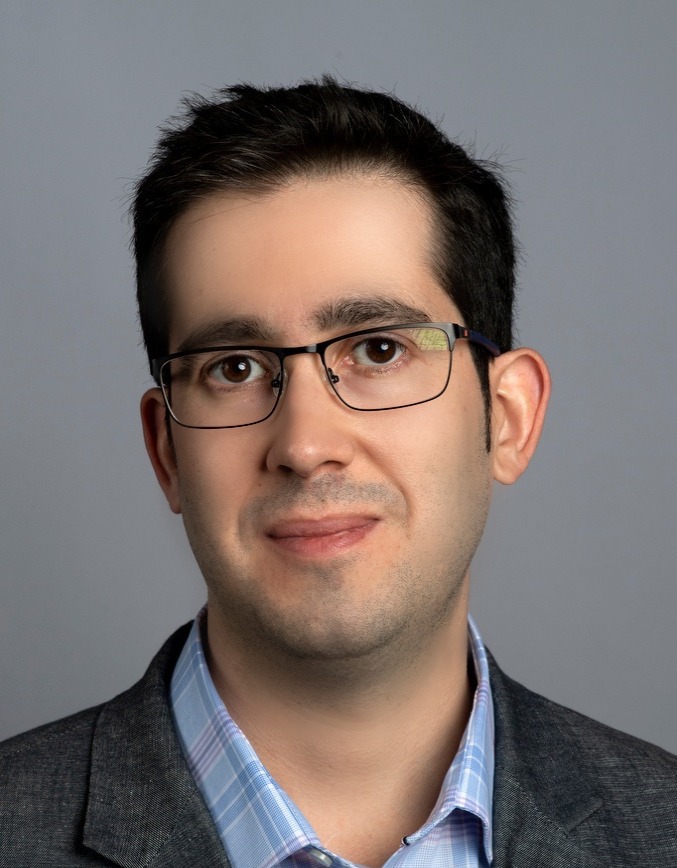}}]{Josep Miquel Jornet} (M'13, SM'20) is an Associate Professor in the Department of Electrical and Computer Engineering, the director of the Ultrabroadband Nanonetworking (UN) Laboratory, and a member of the Institute for the Wireless Internet of Things and the SMART Center at Northeastern University (NU). He received a Degree in Telecommunication Engineering and a Master of Science in Information and Communication Technologies from the Universitat Politècnica de Catalunya, Spain, in 2008. He received the Ph.D. degree in Electrical and Computer Engineering from the Georgia Institute of Technology, Atlanta, GA, in August 2013. His core research is in terahertz communications, in addition to wireless nano-bio-communication networks and the Internet of Nano-Things. In these areas, he has co-authored more than 220 peer-reviewed scientific publications, including 1 book and 5 US patents. His work has received more than 13,800 citations (h-index of 54 as of December 2022). He is serving as the lead PI on multiple grants from U.S. federal agencies including the National Science Foundation, the Air Force Office of Scientific Research and the Air Force Research Laboratory as well as industry. He is the recipient of multiple awards from IEEE and ACM as well as four best paper awards. He is a senior member of the IEEE and an IEEE ComSoc Distinguished Lecturer (Class of 2022-2023). He is also the Editor in Chief of the Elsevier Nano Communication Networks journal and Editor for IEEE Transactions on Communications.\end{IEEEbiography}

\end{document}